\documentclass[traditabstract]{aa} 
\usepackage{graphicx}
\usepackage{txfonts}
\usepackage{longtable}
\begin{document}
   \title{The GALEX Ultraviolet Virgo Cluster Survey (GUViCS). VI: The UV luminosity function of the Virgo cluster and its
   surrounding regions\thanks{Table A.1 
   is available in electronic form at the CDS via anonymous ftp to cdsarc.u-strasbg.fr(130.79.128.5) or via http://cdsweb.u-strasbg.fr/cgi-bin/qcat?J/A+A/}}
   \subtitle{}
  \author{A. Boselli\inst{1},         
	  S. Boissier\inst{1},	  
	  E. Voyer\inst{1},	  
	  L. Ferrarese\inst{2},
	  G. Consolandi\inst{3},	  
	  L. Cortese\inst{4},	
	  P. C{\^o}t{\'e}\inst{2},
	  J.C. Cuillandre\inst{5},  
	  G. Gavazzi\inst{3},
	  S. Gwyn\inst{2},	  
	  S. Heinis\inst{6},	  
	  O. Ilbert\inst{1},
	  L. MacArthur\inst{7},	  
	  Y. Roehlly\inst{1}
       }

\institute{	
		Aix Marseille Universit\'e, CNRS, LAM (Laboratoire d'Astrophysique de Marseille), UMR 7326, F-13388, Marseille, France
             \email{alessandro.boselli@lam.fr, samuel.boissier@lam.fr, elysse.voyer@gmail.com, olivier.ilbert@lam.fr, yannick.roehlly@lam.fr}
         \and  
		NRC Herzberg Astronomy and Astrophysics, 5071 West Saanich Road, Victoria, BC, V9E 2E7, Canada
		\email{laura.ferrarese@nrc-cnrc.gr.ca, patrick.cote@nrc-cnrc.gr.ca, stephen.gwyn@nrc-cnrc.gr.ca}
	\and
	 	Universit\'a di Milano-Bicocca, piazza della scienza 3, 20100, Milano, Italy
		\email{guido.consolandi@mib.infn.it, giuseppe.gavazzi@mib.infn.it}
	\and	
	 	Centre for Astrophysics \ Supercomputing, Swinburne University of Technology, Mail H29-PO Box 218, Hawthorn VIC 3122, Australia
		\email{lcortese@swin.edu.au}
	\and
		CEA/IRFU/SAP, Laboratoire AIM Paris-Saclay, CNRS/INSU, Université Paris Diderot, Observatoire de Paris, PSL Research University, F-91191 Gif-sur-Yvette Cedex, France
		\email{jc.cuillandre@cea.fr}
	\and
	 	Department of Astronomy, University of Maryland, Stadium Drive, College Park, MD 20742-2421, USA
		\email{sebastienheinis@gmal.com}
	\and
		Department of Astrophysical Sciences, Princeton University, Princeton, NJ 08544, USA
		\email{lauren@astro.princeton.edu}
	 }

\authorrunning{Boselli et al.}
\titlerunning{GUViCS: the UV luminosity function of the Virgo cluster}

   \date{}

 
  \abstract  
{ We use the GALEX data of the GUViCS survey to construct the NUV luminosity function of the Virgo cluster over $\sim$ 300 deg.$^2$, an area covering
the cluster and its surrounding regions up to $\sim$ 1.8 virial radii. The NUV luminosity function is also determined for galaxies of different morphological type and
$NUV-i$ colour, and for the different substructures within the cluster. These luminosity functions are robust vs. statistical corrections since based 
on a sample of 833 galaxies mainly identified as cluster members with spectroscopic redshift (808) or high-quality optical scaling relations (10).
We fit these luminosity functions with a Schechter function, and compare the 
fitted parameters with those determined for other nearby clusters and for the field. The faint end slope of the Virgo NUV luminosity function ($\alpha$ = -1.19),
here sampled down to $\sim$ $NUV$ = -11.5 mag,
is significantly flatter than the one measured in other nearby clusters and similar to the field one. Similarly $M^*$ = -17.56 is one-to-two magnitudes fainter
than measured in Coma, A1367, the Shapley supercluster, and the field. These differences seem due to the quite uncertain statistical corrections
and the small range in absolute magnitude sampled in these clusters.
We do not observe strong systematic differences in the overall NUV luminosity function of the core of the cluster 
with respect to that of its periphery. We notice, however, that the relative contribution of red-to-blue galaxies at the faint end is inverted, with red quiescent
objects dominating the core of the cluster and star forming galaxies dominating beyond one virial radius. These observational evidences are discussed in the framework of galaxy
evolution in dense environments.

 }
   {}
   {}
   {}
   {}
   {}


   \maketitle
%

\section{Introduction}

The environment plays a fundamental role in shaping galaxy evolution. Since the seminal work of A. Dressler in the eighties 
it became evident that galaxies in rich environments are systematically different than those located in the field. They are generally quiescent 
ellipticals and lenticulars, in contrast to the dominant population of star forming disc galaxies along the cosmic web 
(Dressler 1980; Binggeli et al. 1988; Whitmore et al. 1993; Dressler et al. 1997). At the same time it became clear that the physical properties of 
the star forming systems inhabiting rich environments such as clusters and groups are also systematically different than those of their isolated analogues, 
with a reduced gas content and activity of star formation (e.g. Boselli \& Gavazzi 2006).\\
Several physical processes have been proposed to explain the origin of these differences. They can be divided in two main families, those 
related to the gravitational interactions between galaxies or with the potential well of the cluster (tidal interactions - Merritt 1983; Byrd \& Valtonen 1990, 
harassment - Moore et al. 1998), and those exerted by the hot and dense intracluster medium on galaxies moving at high velocity within the overdense 
region (ram pressure stripping - Gunn \& Gott 1972, viscous stripping - Nulsen 1982, thermal evaporation - Cowie \& Songaila 1977, starvation - Larson et al. 1980). 
These processes might start to become active well before galaxies entered rich clusters because
these large dynamically bounded structures have been formed through the accretion of smaller groups of galaxies (pre-processing; Dressler 2004).\\
The identification of the dominant environmental process affecting galaxy evolution can be done through the statistical comparison of the 
physical properties of galaxies in a wide range of densities, from the core of rich clusters, compact and loose groups, down to the field, cutting the studied samples in
slice of redshift to trace the time-evolution of the effects. 
Indeed, these processes act on different timescales and are characterised by efficiencies that depend on the density of galaxies and of the intergalactic medium,  
the velocity dispersion within the structure, and the mass of galaxies. The identification of the perturbing process can also be done through the comparison of the
radial and kinematic properties of the 
perturbed galaxies with the predictions of tuned models and simulations.\\

Thanks to its proximity ($\sim$ 17 Mpc, Gavazzi et al. 1999; Mei et al. 2007) and to its unrelaxed structure indicating its still undergoing formation, 
the Virgo cluster has become the ideal target in the local universe for the study
of the effects of the environment on galaxy evolution. The advent of large panoramic detectors made possible the observations of the whole Virgo cluster region 
in the optical bands by the NGVS
survey (Ferrarese et al. 2012), in the 21 cm HI line by ALFALFA (Giovanelli et al. 2005), and in the infrared by \textit{Herschel} (The HeViCS survey, Davies et al. 2010). 
We have recently completed a UV survey of the cluster over $\sim$ 300 deg.$^2$ using the GALEX space observatory (the GUViCS survey; Boselli et al. 2011). 
UV data are of particular interest to trace the content and distribution of the young stellar populations in late-type galaxies 
(e.g. Kennicutt 1998; Boselli et al. 2009), while the one of very evolved stars in quiescent systems (O'Connel 1999; Boselli et al. 2005a).
For these reasons the UV data are critical to trace short-term variations in the star formation activity of late-type systems as those encountered in galaxies
freshly infalling in rich clusters (e.g. Boselli et al. 2006, 2008a,b, 2014a,b; Hughes \& Cortese 2009; Cortese \& Hughes 2009).
The UV data extracted from this survey have been recently published in Voyer et al. (2014), and the study of the physical properties of Virgo cluster galaxies 
derived from a multifrequency analysis in Boselli et al. (2014a). In two other companion papers we presented the study of the extraplanar star formation activity
of a few gas stripped galaxies in the cluster (Boissier et al. 2012), and of the diffuse Galactic emission of the UV scattered light in the Virgo direction (Boissier et
al. 2015). In the present work we derive and study the UV luminosity function of the Virgo cluster and its surrounding regions. This work is thus an extension of the
study presented in Boselli et al. (2011), where the UV luminosity function was derived only for the central 12 deg.$^2$. The luminosity function is a statistical tool
widely used to study the mean properties of large samples of galaxies. Often used in other wavelengths domains, it has been derived at rest frame in the GALEX 
UV bands only for the general field (Wyder et al. 2005) and for a few other nearby clusters (Coma, Cortese et al. 2008; Hammer et al. 2012; A1367, Cortese et al. 2005; 
Shapley, Haines et al. 2011). Compared to these previous works, the UV luminosity function derived from the GUViCS survey has the major advantage of including a significant
larger number of objects with direct redshift measurements, of sampling a much wider range in galaxy density, from the core of the cluster out to the periphery ($\sim$ 2
$R_{vir}$), and of including the dwarf galaxy population ($NUV$ $\lesssim$ -11.5 mag). Dwarf systems are of particular interest in the study of the effects of the environment
on galaxy evolution since they are fragile objects easily perturbed just because of their shallow potential well (e.g. Boselli \& Gavazzi 2014). \\
The paper is structured as follows: in section 2 we describe the dataset used in this analysis, while in section 3 we explain the different techniques used to identify
Virgo cluster members. The UV luminosity function is derived in section 4 and analysed in the framework of galaxy evolution in rich environments in section 5.

\section{The data}

The GUViCS survey has covered the Virgo cluster in the region between 12h $\leq$ $R.A$ $\leq$ 13h and 0$^o$ $\leq$ $dec$ $\leq$ 20$^o$. 
In this region the sky coverage is 94 \% ~ (Boselli et al. 2014a) at the depth of the AIS, where the completeness is 20 magnitude, and  
65 \% ~ at the depth of the MIS (21.5 mag)\footnote{The coverage is less than 100 \% because of the presence of a few bright young stars
which have been expressly avoided during the survey since saturating the UV detector}. 
Unfortunately, because of a failure of the FUV detector, this survey has been completed 
only in the NUV. The analysis presented in this work will thus be limited to this band.
The number of NUV detected sources in the covered region is more than 1.2 millions (Voyer et al. 2014). These are Milky Way stars, nearby galaxies 
and background sources. To construct the NUV luminosity function of the Virgo cluster and of its surrounding regions we have 
to count the number of galaxies in a given volume down to a limiting magnitude. For consistency with our previous works we 
consider as Virgo members those galaxies with a recessional velocity $vel$ $\leq$ 3500 km s$^{-1}$. This limit has been chosen
to include all the objects belonging to the different cluster substructures and their surrounding regions (e.g. Boselli et al. 2014a).
As limiting magnitude we adopt $NUV$=20 mag. Down to this magnitude, the catalogue is 100\% complete over the $\sim$ 300 deg.$^2$ covered by the survey. 

Voyer et al. (2014) provide integrated magnitudes for extended sources (UV-VES catalogue) included in the widely known catalogues of nearby galaxies
(NGC, UGC, IC, VCC, ...), for a total of 1770 galaxies. These magnitudes have been extracted with the FUNTOOLS analysis package on DS9
using manually defined apertures adjusted to fit the full UV profile of each traget galaxy. 
The same procedures have been adopted for all the sources identified as Virgo cluster members using spectroscopic redshifts. These procedures are
necessary for extended sources where the automatic GALEX pipeline often does not recognise the extended nature of the galaxies from the
point-like structure of HII regions, galaxy nuclei or complex high surface brightness features present in star forming systems. 
For the other sources, NUV fluxes have been taken from the catalogue of point-like objects (UV-VPS) of Voyer et al. (2014).
Down to the limiting magnitude of $NUV$=20 mag this catalogue includes 33026 objects. These are generally faint objects, where the NUV magnitudes 
extracted from the pipeline are accurate enough for the purpose of this work. Indeed, as clearly shown in Boselli et al. (2011) (Fig. 7),
the difference between extended magnitudes and those extracted from the standard GALEX pipeline optimised for point-like sources is $\lesssim$ 0.3 mag whenever $NUV$ $\gtrsim$ 17 mag.
UV magnitudes are corrected for Galactic attenuation using the Schlegel et al. (1998) map combined with the Fitzpatrick \& Massa (2007) 
extinction curve. Given the high Galactic latitude of the Virgo cluster, these corrections are low ($A(NUV)$ $\lesssim$ 0.4 mag; Boissier et al. 2015). 

For all galaxies at the distance of Virgo ($vel$ $\leq$ 3500 km s$^{-1}$) we also extracted 22 $\mu$m WISE fluxes necessary for the dust
internal attenuation corrections. These fluxes have been extracted from the original WISE fits images with the same procedures used for the UV data, 
consistently with Boselli et al. (2014a). 
The dust extinction correction is done following the prescription of Hao et al. (2011).
NUV corrected data are used to estimate the star formation rate in star forming galaxies using again the prescription of Hao et al. (2011).

Distances are necessary to calculate absolute magnitudes. For consistency with our previous works, distances are determined assuming 
the mean distance of the different cluster subgroups (see Table 3 in Boselli et al. 2014a). Galaxies at the periphery of the cluster 
are all assumed at 17 Mpc, the distance of the main body of Virgo.

\section{The identification of Virgo members}

The construction of the NUV luminosity function needs first the identification, among the UV detections with $NUV$ $\leq$ 20 mag, 
of all galaxies at the distance of Virgo. This is done using spectroscopic redshift whenever available (see sect. 3.1), while photometric
relations (sect. 3.2) or statistical considerations otherwise (sect. 3.3). 

\subsection{Identification based on spectroscopic data}

Voyer et al. (2014) has cross-matched all the UV detections with the spectroscopic catalogues available on the net 
(NED\footnote{https://ned.ipac.caltech.edu/}, SDSS DR9\footnote{https://www.sdss3.org/dr9/}). 
For these objects the spectroscopic information 
can be easily used to identify Virgo members ($vel$ $\leq$ 3500 km s$^{-1}$). Out of the 1770 galaxies listed in the extended source catalogue, 
688 galaxies match these criteria. There are, however, 34 extended galaxies with $NUV$ $\leq$ 20 mag 
but without any redshift information. Among these, we identify the possible cluster members using the same procedures adopted for the other 
galaxies included in the UV point source catalogue (see below).

As previously mentioned, down to $NUV$ $\leq$ 20 mag the majority of the point-like detected objects are not Virgo cluster members but
rather Galactic stars and background sources. Given the poor angular resolution of the GALEX images ($\sim$ 4-5 arcsec), the separation
of point-like stars from extended galaxies requires the use of optical images of better quality.
Considering the typical UV-optical colour magnitude relation of galaxies in the Virgo cluster (e.g. Boselli et al. 2014a), a cut at $NUV$ = 20 mag 
roughly corresponds to a cut in the SDSS $g$ band of $\sim$ 19 mag for the blue star forming objects and $g$ $\sim$ 16 mag for the 
quiescent dwarf ellipticals. At this depth the SDSS is complete both in terms of total magnitude and surface brightness  
and can thus be used for this purpose. Indeed, out of the 33026 GUViCS point-like sources with $NUV$ $\leq$ 20 mag, 32277 (98\%) 
have an optical counterpart in the SDSS (the 749 objects without a SDSS counterpart are mainly ghosts of saturated stars, HII regions in resolved galaxies, or
multiple systems unresolved in the UV images). About one third of these (7532) are photometrically identified as galaxies in the SDSS (SDSS\_phot\_ObjType=3). 
We have cross-matched these 7532 objects with the SDSS DR12 spectroscopic catalogue, with NED, and with eBOSS through the DR12 SDSS finding chart tool. 
Spectra are available for 5620 galaxies (75\% of the sample). 
Their optical image on the SDSS DR12 finding chart revealed, however, that a significant fraction of them are 
bright stars. Generally saturated, they are extended on the SDSS image and thus missclassified as galaxies by the automatic SDSS pipeline. 
Among the bona fide galaxies of the UV-VPS catalogue of Voyer et al. (2014) with available redshift we identify 125 objects with $vel$ $\leq$ 3500 km s$^{-1}$. 
Five of these, however, are left out
of the sample because their NUV magnitude determined using our ad-hoc procedures for extended sources are below the limiting value of $NUV$ = 20 mag.
The resulting sample of Virgo cluster members determined using spectroscopic data consists thus of 808 galaxies. 

\subsection{Identification using optical NGVS scaling relations}
 
The identification of Virgo members among those objects without any spectroscopic data is critical in the determination of the luminosity function.
The membership criteria adopted by the NGVS, fully described in Ferrarese et al. (2015), give the probability that a galaxy belongs to Virgo according to its location 
in a multi-parameter space defined by a combination of galaxy structural parameters, photometric redshifts (Raichoor et al. 2014), and an index measuring the strength of residual 
structures in images created by subtracting from each galaxy a best fitting Sersic model. The exact combination of axes in this
space was selected to allow for maximum separation of known (mostly spectroscopically confirmed) Virgo members and background sources, 
the latter identified in control fields located at three virial radii from the cluster centre. The NGVS covers 104 square degrees, and 
its entire footprint is mapped by GUViCS.
We thus cross-matched the GUViCS UV-VPS catalogue with the NGVS catalogue of Virgo members identified using
their photometric parameters and found 10 UV sources identified as Virgo members in the NGVS.
We assume all these objects at a distance of 17 Mpc, the typical distance of Virgo and of its
surrounding regions. The depth of the NGVS survey, which is 3 mag deeper than the SDSS in terms of surface brightness sensitivity, 
guarantees that all UV detected galaxies down to NUV $\leq$ 20 mag have an optical counterpart.

\subsection{Identification using statistical estimates oustide the NGVS footprint}

The NGVS covers Virgo out to one virial radius, for a total of 104 square degrees. To identify Virgo members in the GUViCS areas beyond the NGVS 
boundaries, and in the absence of spectroscopic information, we must therefore resort to statistical arguments.
In the GUViCS UV-VPS catalogue there are 1884 UV galaxies without any membership information. We drop those located 
within the NGVS footprint, reducing the sample of objects with no available membership information to 1155 sources.
We then use three different statistical methods to estimate the number of potential members among these galaxies. 

\subsubsection{Statistical correction based on NGVS results}

If 10 galaxies with no redshift in the NGVS footprint (104 deg.$^2$) are identified as Virgo members 
using criteria adopted by the NGVS, and the density of galaxies is conserved over the surveyed region, 20 extra objects are expected 
to be present in the remaining $\sim$ 200 deg.$^2$ covered by GUViCS. Since the NGVS observed the central region of the cluster, 
where the density is at its maximum, this number should be considered as an upper limit.

\subsubsection{Statistical correction based on redshift measurements}

Out of the 5620 galaxies in the VPS catalogue of Voyer et al. (2014) with spectra in the 300  deg.$^2$ of the GUViCS survey only 125
have $vel$ $\leq$ 3500 km s$^{-1}$. If we assume that the same ratio of members/background galaxies is conserved among those objects
without any spectroscopic information (1884), we would thus expect 42 new members. This number drops to 30 if we limit this comparison
in the region located outside the NGVS footprint corresponding to the cluster periphery (90/3452). 

\subsubsection{Statistical correction based on GALEX-SDSS scaling relations}

\begin{figure*}
   \centering
   \includegraphics[width=17cm]{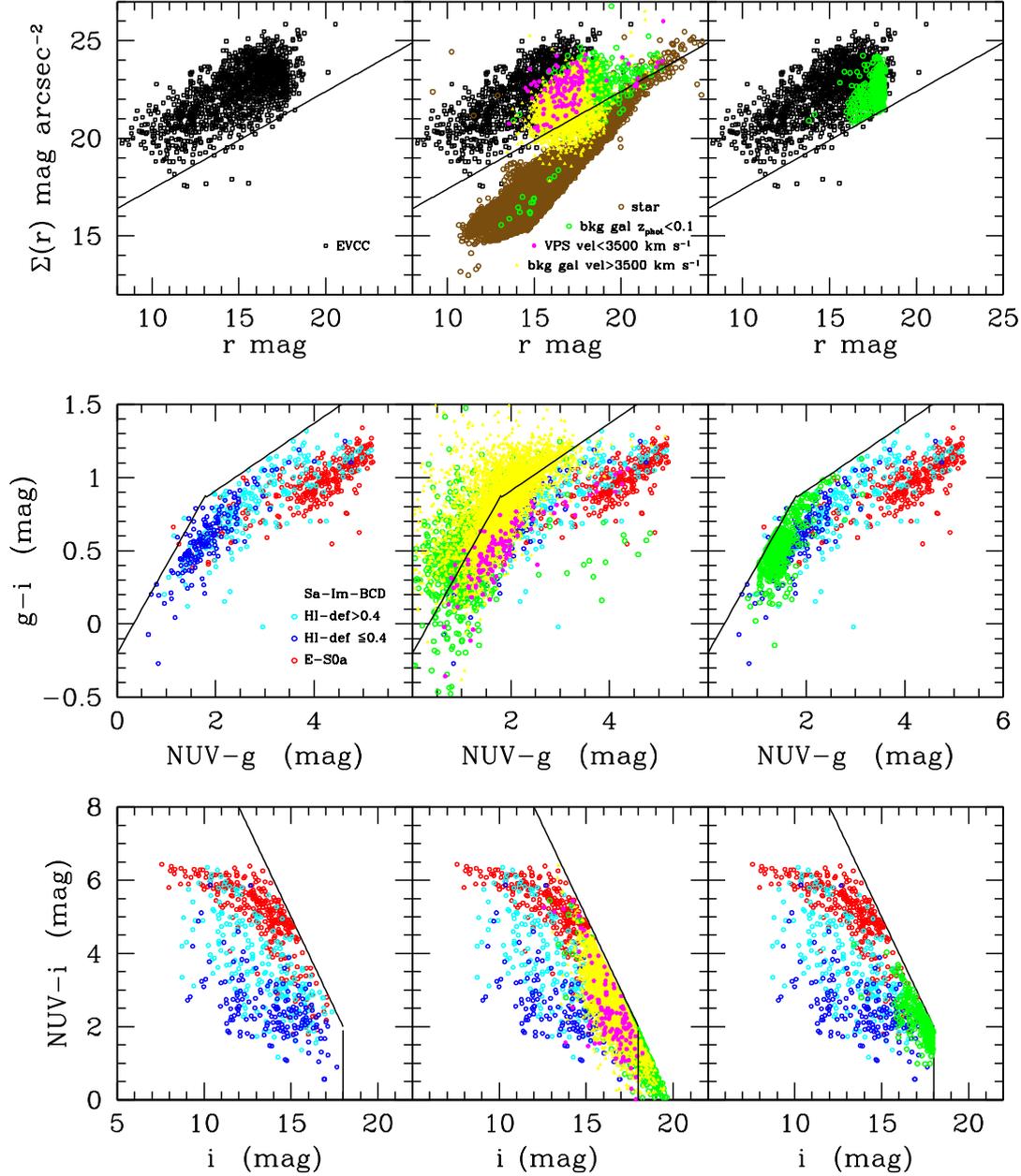}
   \caption{The $\Sigma (r)$ vs. $r$ (upper row), $g-i$ vs. $NUV-g$ (middle row), and $NUV-i$ vs. $i$ (lower row)
   scaling relations used to identify Virgo cluster possible members. Left panels: scaling relations of
   spectroscopically confirmed Virgo members included in the extended source GUViCS catalogue. The black empty squares 
   indicate the local EVCC galaxies. The red, cyan, and blue empty circles in the middle and lower panels indicate the
   early-type and late-type galaxies, these last selected according to their HI-gas content (HI-def$\leq$ 0.4, blue symbols; 
   HI-def$>$ 0.4, cyan symbols). The black solid lines 
   indicates the limits in the different scaling relations used to identify possible Virgo cluster members.
   Central panels: relations drawn when all optically identified galaxies with $NUV$ $<$ 20 mag from the GUViCS VPS catalogue are included.
   Stars, plotted only in the central upper panel, are indicated with a brown open circle.
   VPS galaxies with $vel$ $\leq$ 3500 km s$^{-1}$ are indicated with magenta filled dots, spectroscopically confirmed
   background sources with filled yellow triangles, while potential members with photometric redshift $\leq$ 0.1 with green
   empty circles. Right panels: scaling relations of spectroscopically confirmed Virgo members (as in the left panels) 
   with candidate members from the VPS catalogue selected using the criteria given in sect. 3.3.3 (green empty circles).
   All these relations are determined using UV data corrected for Galactic dust attenuation.
 }
   \label{sigmaall}%
   \end{figure*}

An alternative method for identifying possible members has been first proposed by Boselli et al. (2011). For the same reasons given above, 
cluster galaxies populate different regions in the optical and UV surface brightness vs.
observed magnitude relations and can thus be discriminated using their photometric parameters. Unfortunately the relations proposed by 
Boselli et al. (2011) are not sufficiently accurate to identify Virgo members among quiescent dwarf galaxies such as those located
in the central 12 deg.$^2$ analysed in that work. Here we propose an alternative technique based on three different scaling relations ($\Sigma (r)$ vs. $r$
mag; $g-i$ vs. $NUV-g$, $NUV-i$ vs. $i$ mag, where $\Sigma(r)$ is the mean surface brightness within the half-light
radius\footnote{The GUViCS catalogue lists the Petrosian half-light radius, while the EVCC the Kron half-light radius. We have checked that, 
for galaxies belonging to both samples, these radii are consistent.})\footnote{All values are corrected for Galactic extinction. These relations are also different from those
adopted by Ferrarese et al. (2015) based only on optical data.} 
which we demonstrate to be fairly efficient at rejecting a significant number of background objects 
also among star forming systems such as those detected in the outer regions of the cluster (see Fig. \ref{sigmaall}). 
Despite their relative low scatter, these relations are unfortunately not sufficiently accurate for an unambiguous identification of cluster members 
and thus need to be combined with statistical tests.

These scaling relations are determined using either the Extended Virgo Cluster Catalogue (EVCC, Kim et al. 2014) 
or the GUViCS catalogue of galaxies analysed in Boselli et al. (2014a).
The EVCC catalogue has been chosen because it includes galaxies at the same distance as Virgo ($vel$ $\leq$ 3000 km s$^{-1}$), it has both star forming
and quiescent galaxies, and provides structural parameters comparable to those listed in the GUViCS point-like catalogue of Voyer et al. (2014). The EVCC 
catalogue is used to determine the $\Sigma(r)$ vs. $r$ apparent magnitude relation. 
Unfortunately the EVCC catalogue does not include UV magnitudes and thus cannot be used in the $g-i$ vs. $NUV-i$ colour-colour and the 
$NUV-i$ vs. $i$ relations. For these UV based relations we use as comparison sample the one composed of Virgo cluster galaxies analysed in Boselli et
al. (2014a). Galaxies are separated according to their morphological type (early- from late-type) and atomic gas content. 
For consistency, the Virgo cluster sample is limited to $NUV$ $\leq$ 20 mag. 

Figure \ref{sigmaall} shows that: \\
a) a number of sources identified by the standard SDSS pipeline or by our visual inspection of the SDSS finding charts as galaxies
are saturated stars. They have $r$-band surface brightnesses $\sim$ 5 mag brighter than those of typical galaxies of similar magnitude, but
are rather comparable to those of stars. We identify as possible members in the $\Sigma (r)$ vs. $r$ mag plot all galaxies satisfying
the condition $\Sigma (r)$ $>$ 0.5$\times$$r$+12.4. We recall that this criterion is different and less strict than 
the one used in Boselli et al. (2014a) (the number of rejected objects is relatively small). 
This difference is due to the fact that in the inner 12 deg.$^2$ 
at faint magnitudes the population of Virgo members is strongly dominated by dwarf ellipticals with red colours. Thus the difference 
in $\Sigma (r)$ with background star forming galaxies was more pronounced than in the present sample, which includes
local star forming systems at the periphery of the cluster.\\
b) all Virgo cluster members avoid the region above the $g-i$ = 0.6 $\times$ $(NUV-g)$ - 0.2 and $g-i$ = 0.23 $\times$ $(NUV-g)$ + 0.45 relations
in the $g-i$ vs. $NUV-g$ colour-colour diagram. These regions are populated mainly by background sources. \\
c) the sharp cut in the $NUV-i$ vs. $i$ colour-magnitude relation is due to the adopted cut in NUV. This relation, however, shows that
all identified members have observed $i$ magnitudes $\leq$ 18.\\

These relations can be combined with a limit on the photometric redshift, accurately determined on these bright sources, to reject background galaxies. 
This is done by removing all objects with $z_{photNN}$ $\geq$ 0.1, where $NN$
stands for the Neural Network (NN) photometric redshift definition (D'Abrusco et al. 2007). This cut in photometric redshift, which corresponds
to the one adopted by Boselli et al. (2011) for the rejection of background galaxies in the inner 12 deg.$^2$ of the cluster, has been chosen 
to maximise the number of background rejected objects and minimise the exclusion of possible members. With this new selection criteria, the sample 
of possible members outside the NGVS footprint drops to 264 objects. The number of spectroscopically confirmed Virgo galaxies satisfying the same
photometric criteria is 75/90 (83.3\%), while that of members over background objects 75/1636 (4.6\%). We thus expect that 15 out of the 264 galaxies 
satisfying the criteria are at the distance of Virgo. 

\subsubsection{The applied statistical correction}

\begin{figure}
   \centering
   \includegraphics[width=15cm]{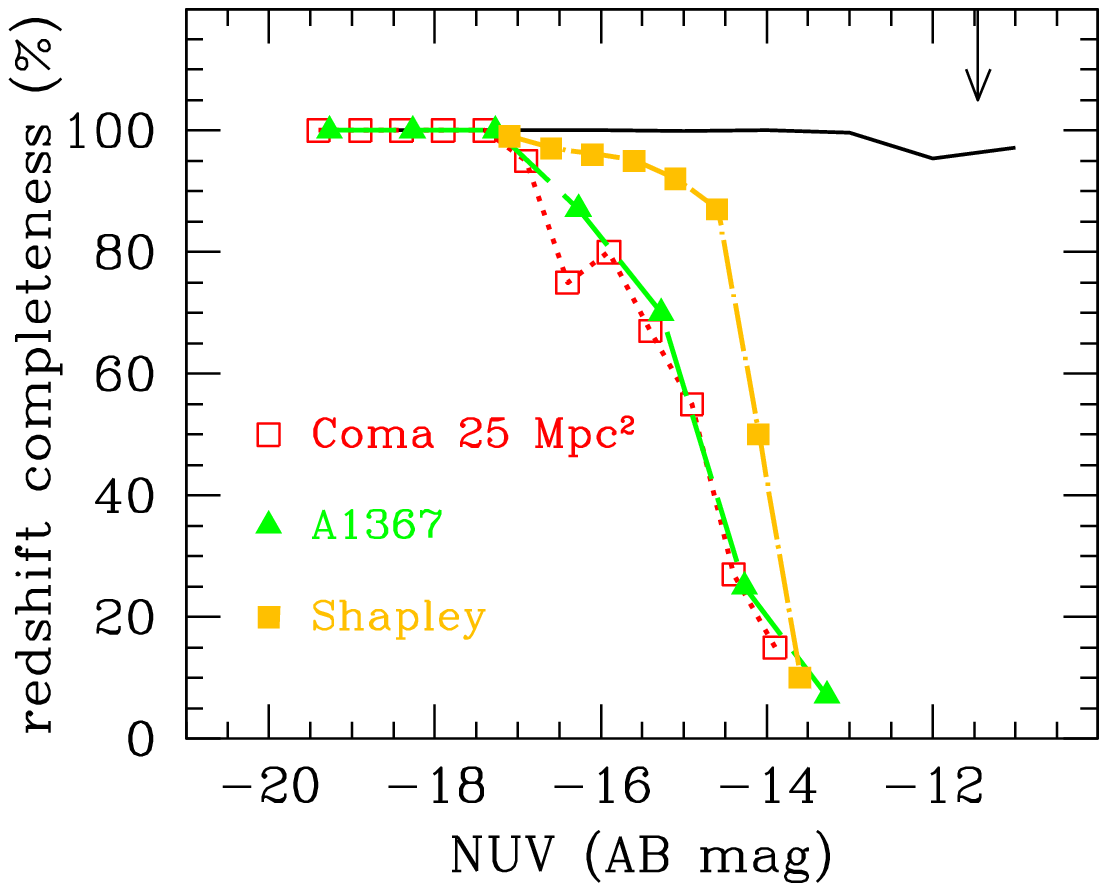}
   \caption{The NUV redshift completeness of the GUViCS sample compared to that of Coma (red open squares; Cortese et al. 2008),
   A1367 (green filled triangles; Cortese et al. 2005), and Shapley supercluster (orange filled squares; Haines et al. 2011).
   The vertical arrow shows the photometric completeness limit of the present survey ($NUV$ = -11.45).
 }
   \label{histoNUV}%
   \end{figure}

The three statistical methods described above give consistent results. We thus include 15 extra objects to those already selected 
spectroscopically (see sect. 3.1) or using NGVS data (sect. 3.2) in the determination of the NUV luminosity function. 
The distribution of the 264 galaxies selected according to 
Fig. \ref{sigmaall} is used to make the appropriate statistical correction
as a function of the NUV absolute magnitude in the determination of the NUV luminosity function (see Fig. \ref{histoNUV}). These 15 
galaxies are all assumed to be at a distance of 17 Mpc. Obviously the correction is more important at faint luminosities. Figure \ref{sigmaall}
also shows that these statistically added galaxies, all located in the outer regions of the cluster, have blue colours.

The resulting sample is thus composed of 808 spectroscopically confirmed sources, 10 objects identified at the distance of the Virgo cluster using 
NGVS based scaling relations, and 15 extra objects added for statistical considerations, for a total of 833 galaxies.
This indicates that the determination of the NUV luminosity function 
is very robust against statistical uncertainties, making the Virgo cluster the ideal target for such studies in the nearby universe
(e.g. Rines \& Geller 2008).

\subsection{Possible biases}

The identification of galaxies through automatic pipelines is first done on surface brightness criteria. 
Low surface brightness, extended systems are hardly detectable even if their integrated magnitude is above
the limiting magnitude requested for the completeness of the survey. These systems might thus be potentially missed 
in the GUViCS survey. This effect, however, should be limited because 65\% ~ of the Virgo cluster region has been mapped at the depth of the 
GALEX MIS ($\sim$ 1500 sec), where the completeness limit of 22.7 mag (Morrissey et al. 2007) is significantly deeper than the one adopted in this work.
At this depth low surface brightness features can be detected down to $\sim$ 28.5 mag arcsec$^{-2}$ (e.g. Boselli et al. 2005b).
Furthermore, Voyer et al. (2014) have systematically checked the UV images of all the optically selected VCC galaxies. The analysis presented in 
Boselli et al. (2011) revealed that most of the UV detected galaxies used to construct the UV luminosity function down to $NUV$ = 21 mag 
in the central 12 deg.$^2$ of Virgo (1 mag below the UV limit adopted in this work) have a counterpart in the VCC catalogue. This central region, however, is dominated by
quiescent galaxies with red colours, thus bright in the optical bands if detected in the NUV. 

None of the three extremely low surface brightness extended galaxies ($\Sigma (g)$ $\sim$ 28 mag arcsec$^{-2}$) recently discovered by Mihos et al. (2015) in the core of the Virgo 
cluster has been detected by GALEX. Given their total magnitude ($g$ $\gtrsim$ 17 mag), and the typical colour of quiescent dwarf ellipticals ($NUV-g$ $\gtrsim$ 3 mag), we would expect that 
their total NUV magnitude is $\gtrsim$ 20 mag, thus below the limiting magnitude adopted for the determination of the UV luminosity function. Less extreme objects such as those observed in Coma by
van Dokkum et al. (2015) and Koda et al. (2015) are easily detected by the NGVS and are thus considered in the NUV vs. optical cross-identification.
The NGVS footprint largely overlaps with the VCC region, and has been observed in four photometric bands ($ugiz$) at a point-source depth of $g$ $\sim$ 25.9 mag 
and a surface brightness limit $\Sigma(g)$ $\sim$ 29 mag arcsec$^{-2}$ (Ferrarese et al. 2012). This limit in surface brightness is roughly 4 mag below the one reached by the VCC 
(Ferrarese et al. 2012). Among the UV detected galaxies with $NUV$ $<$ 20 mag, 43 objects identified as Virgo members from spectroscopic measurements or using 
the NGVS scaling relations have an optical counterpart in the NGVS 
but not in the VCC. Outside the NGVS footprint optical data are available from the SDSS. Its sensitivity in surface brightness is $\sim$ 26 mag arcsec$^{-2}$, thus
$\sim$ one mag better than the VCC. At the depth of the present survey ($NUV$ $<$ 20 mag) most of the UV detected sources have a counterpart in the SDSS (98\%, see previous
section), it is thus likely that the number of Virgo cluster galaxies without an optical counterpart missed in the present analysis is low. A rough estimate of this
number can be derived thanks to the NGVS. If the density of galaxies is conserved all over the GALEX surveyed region, and if we assume that the sensitivity of the VCC
is comparable to that of the SDSS, we expect to miss 86 galaxies (2 $\times$ 43). Once again this number should be considered as an upper limit because i) the density of
galaxies significantly drops outside the central regions of the cluster, and ii) the SDSS is $\sim$ 1 mag in surface brightness more sensitive than the VCC. 
Furthermore, this number is comparable to the number of galaxies statistically added in the previous section (15).
We thus expect that the number of faint galaxies missed for their low surface brightness in UV is small compared to the total number of galaxies.

Another possible source of uncertainty is the presence of extremely compact extragalactic sources erroneously identified as stars in the SDSS optical images
whose typical resolution is of the order of $\sim$ 1 arcsec. 
Typical examples are the ultra compact dwarf galaxies (UCDs), a population of extragalactic sources with characteristics in between those of compact elliptical galaxies and
globular clusters (Hilker et al. 1999; Drinkwater et al. 2000; Zhang et al. 2015). These sources, however, have NUV apparent magnitudes below the $NUV$ = 20 mag limit because of their relatively red colours 
(Boselli et al. 2011) and should not be counted in the determination of the NUV luminosity function of Virgo. Blue compact dwarf galaxies (BCD), characterised by an intense
activity of star formation, have angular dimensions of $\gtrsim$ 10 arcsec at the distance of Virgo and are thus fully resolved by the SDSS. They are thus included in the
present analysis.

\section{The NUV luminosity function}

\subsection{The NUV luminosity function of the whole cluster}

   \begin{figure}
   \centering
   \includegraphics[width=15cm]{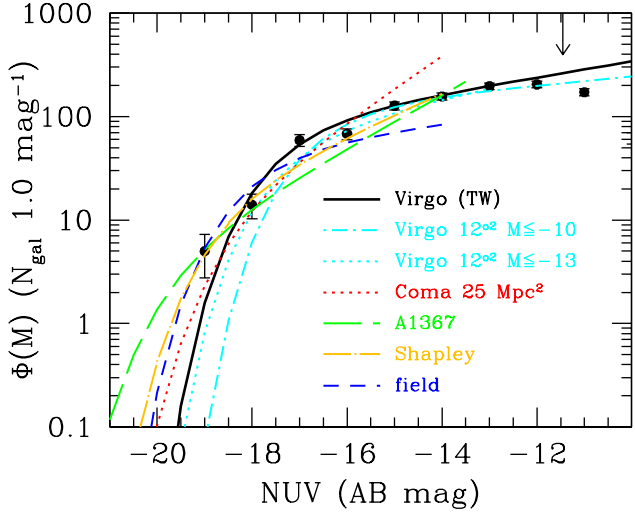}
   \caption{The NUV luminosity function of the Virgo cluster (filled dots). The best fit Schechter function (black solid line)
   is compared to the one determined for the central 12 deg.$^2$. of the cluster by Boselli et al. (2011) at a limiting 
   absolute magnitude of NUV $\leq$ -10 (cyan dotted-dashed line) and  NUV $\leq$ -13 (cyan dotted line), and to that of the Coma cluster
   (red dotted line, from Cortese et al. 2008), A1367 (green long dashed line, Cortese et al. 2005), Shapley supercluster (orange dotted-dashed line; Haines et al. 2011)
   and the field (blue dashed
   line, Wyder et al. 2005). The different luminosity functions have been normalised to $\sim$ the same number of objects above the completeness limit of each cluster for a
   fair comparison. The vertical arrow shows the completeness limit of the present survey ($NUV$ = -11.45).}
   \label{LFFT}%
   \end{figure}

\begin{figure}
   \centering
   \includegraphics[width=10cm]{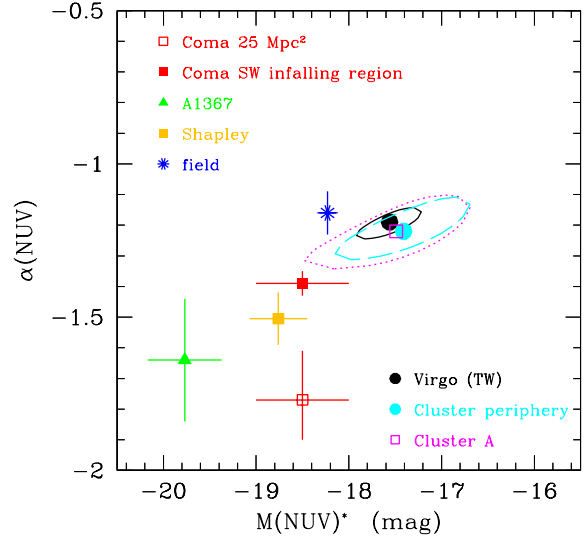}
   \caption{Comparison of the parameters obtained after fitting a Schechter function to the NUV luminosity function of the Virgo cluster (black open circle),
   compared to those determined in Coma 
   over 25 Mpc$^o$ $^2$ (red empty square, Cortese et al. 2008) or in its SW infalling region (red filled square, Hammer et al. 2012), 
   in A1367 (green filled triangle, Cortese et al. 2005), in the Shapley supercluster (orange filled square, Haines et al. 2011),
   and the field (blue asterisk, Wyder et al. 2005). Contours indicate 1 $\sigma$ probability distribution of the two correlated parameters. }
   \label{LFpara}
   \end{figure}

Figure \ref{LFFT} shows the NUV luminosity function of the whole cluster determined over the 300 deg.$^2$ of the GUViCS survey. 
The number of galaxies per bin of absolute NUV magnitude\footnote{Corrected for Galactic attenuation.} are given in Table \ref{LFtab}.
Given that the virial radius $R_V$ of cluster A, the main substructure composing Virgo, is 5.38 degrees (McLaughlin 1999; Ferrarese et al. 2012,
corresponding to 1.60 Mpc at the distance of 17 Mpc), 
this luminosity function is representative of the cluster up to $\sim$ 1.8 virial radii. For comparison with other clusters 
we recall that the core radius of Virgo is of 0.45$^o$ (130 kpc; Boselli \& Gavazzi 2006).
We fit the data with a Schechter function with parameters given in Table \ref{LFfit}. Following Ilbert et al. (2005) we use the MINUIT
package of the CERN library (James \& Roos 1995) to minimise the likelihood (MIGRAD procedure), to obtain the non-parabolic error for each parameter (MINOS procedure), 
and the error contour $\alpha$ - $M^*$ (MNCONT procedure). 
The parameters of the fit are compared to those determined
for the central 12 deg.$^2$ of Virgo by Boselli et al. (2011), for other clusters and for the field in Fig. \ref{LFpara}.

The NUV luminosity function of Virgo is well fitted by a Schechter function with $M^*$ = -17.56 and $\alpha$ = -1.19.
These values are close to those determined in the central 12 deg.$^2$ by Boselli et al. (2011), but are significantly
different from those determined in Coma (Cortese et al. 2008, Hammer et al. 2012), 
A1367 (Cortese et al. 2005), in the Shapley supercluster (Haines et al. 2011), or in the field (Wyder et al. 2005).
The slope of the faint end of the NUV luminosity function of Virgo ($\alpha$ = -1.19) is significantly flatter
than the one of Coma, A1367, or the Shapley supercluster ($\alpha$ -1.5/-1.6) as first noticed by Boselli et al. (2011). 
The slope is however comparable to the one determined in the field ($\alpha$ = -1.16; Wyder et al. 2005). $M^*$ is $\sim$ 1 mag fainter than in Coma, 
in the Shapley supercluster, and in the field, and 2 mag fainter than in A1367.

\subsection{The luminosity function in the different cluster substructures}

The comparison of the properties of the luminosity function determined in different environments, from the cluster periphery to the densest regions
in the core of the cluster, are useful to constrain statistically the role played by the environment in shaping galaxy evolution. The determination 
of the statistical properties of the different cluster substructures, accreted at different epochs, is important to understand the role of pre-processing
in the transformation occurring in cluster galaxies.

\subsubsection{Cluster A vs. periphery}

\begin{figure}
   \centering
   \includegraphics[width=16cm]{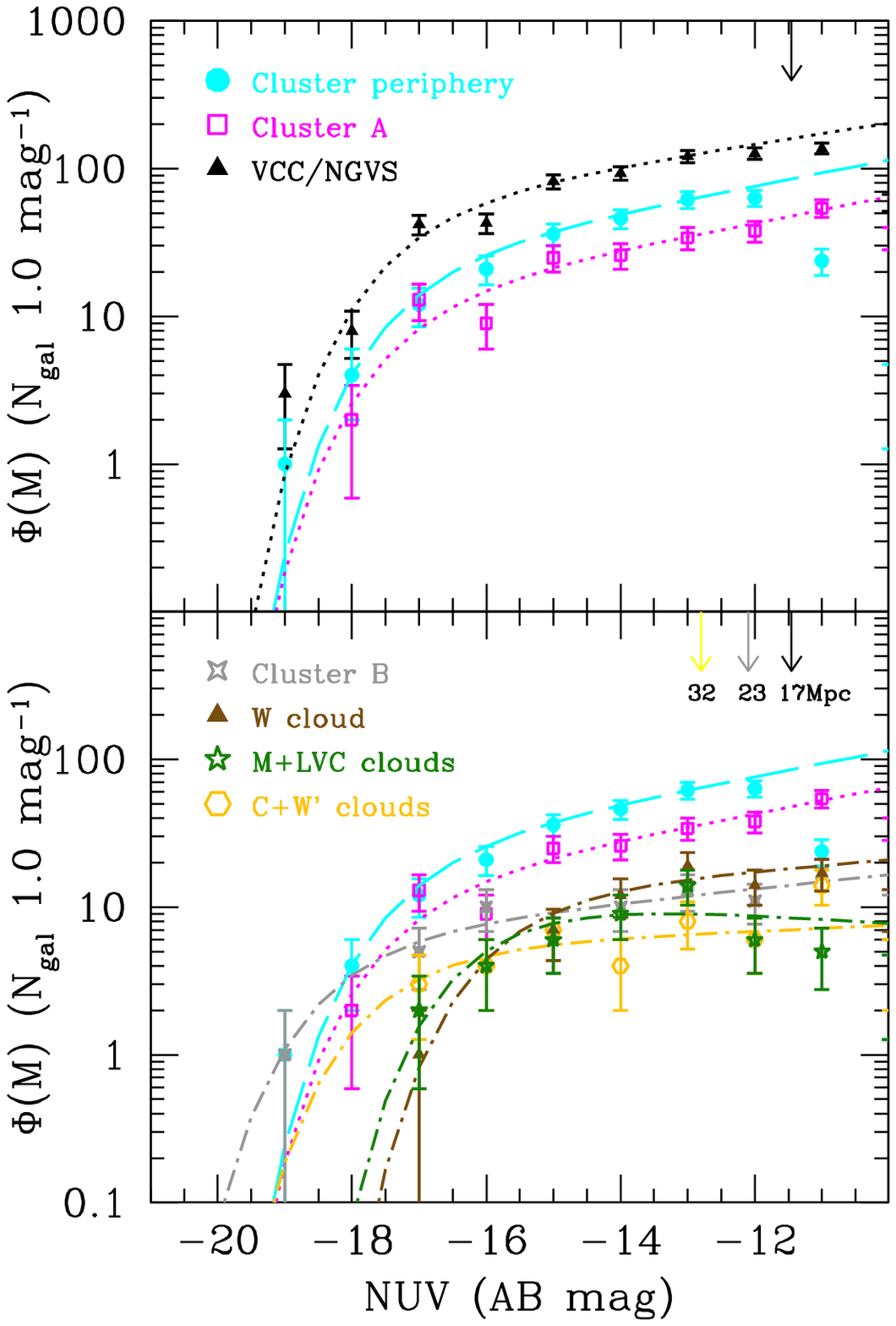}
   \caption{Upper panel: the NUV luminosity function of Virgo 
   cluster A (magenta empty squares) and of the cluster periphery (cyan filled dots) are compared to the one determined
   over the 104$^o$ $^2$ sampled by the VCC and the NGVS survey (black filled triangles). The best fit Schechter functions 
   are shown by the magenta dotted, cyan long dashed, and black dotted lines, respectively. The vertical arrow shows the limit of completeness 
   of the luminosity function. Lower panel: the same cluster A and periphery luminosity functions are compared to those determined for
   cluster B (grey empty stars), W cloud (brown filled triangles), M plus Low Velocity Clouds (dark green empty stars), and cluster C plus W' cloud (orange empty
   hexagons). The vertical arrows show the different limits of completeness for the different cluster substructures: 17 Mpc for cluster A, cluster C, LVC,
   and the cluster periphery, 23 Mpc for cluster B and W' cloud, 32 Mpc for W and M clouds. }
   \label{LFFTall}%
   \end{figure}

Thanks to the large number of objects we can determine the NUV luminosity function separately for the core of the cluster and for its periphery.
Consistently with Boselli et al. (2014a), we select galaxies within cluster A as those objects with $vel$ $<$ 3500 km s$^{-1}$ and with 
an angular distance from M87 $<$ 2.692$^o$, corresponding to half its virial radius. The periphery is composed by all galaxies with $vel$ $<$ 3500 km s$^{-1}$
and an angular distance from M87 $>$ 6.1$^o$ ($>$ 1.1 virial radii from the main body of the cluster).
The NUV luminosity functions of cluster A and of the cluster periphery, whose values are given in Table \ref{LFcloudstab}, 
are fitted with a Schechter function and compared in Fig. \ref{LFFTall}, while the fitted parameters are given in Table \ref{LFfit}. The fitted parameters  
for cluster A and for the periphery
are compared to those determined for the whole cluster and for other clusters in Fig. \ref{LFpara}.

Figures \ref{LFpara} and \ref{LFFTall} and Table \ref{LFfit} show that the NUV luminosity function, as in Coma (Cortese et al. 2008),
does not significantly change from the core of cluster A to the cluster periphery outside $\gtrsim$ 1.1 $R_{vir}$. 

\subsubsection{The other cluster substructures}
   
\begin{figure}
   \centering
   \includegraphics[width=10cm]{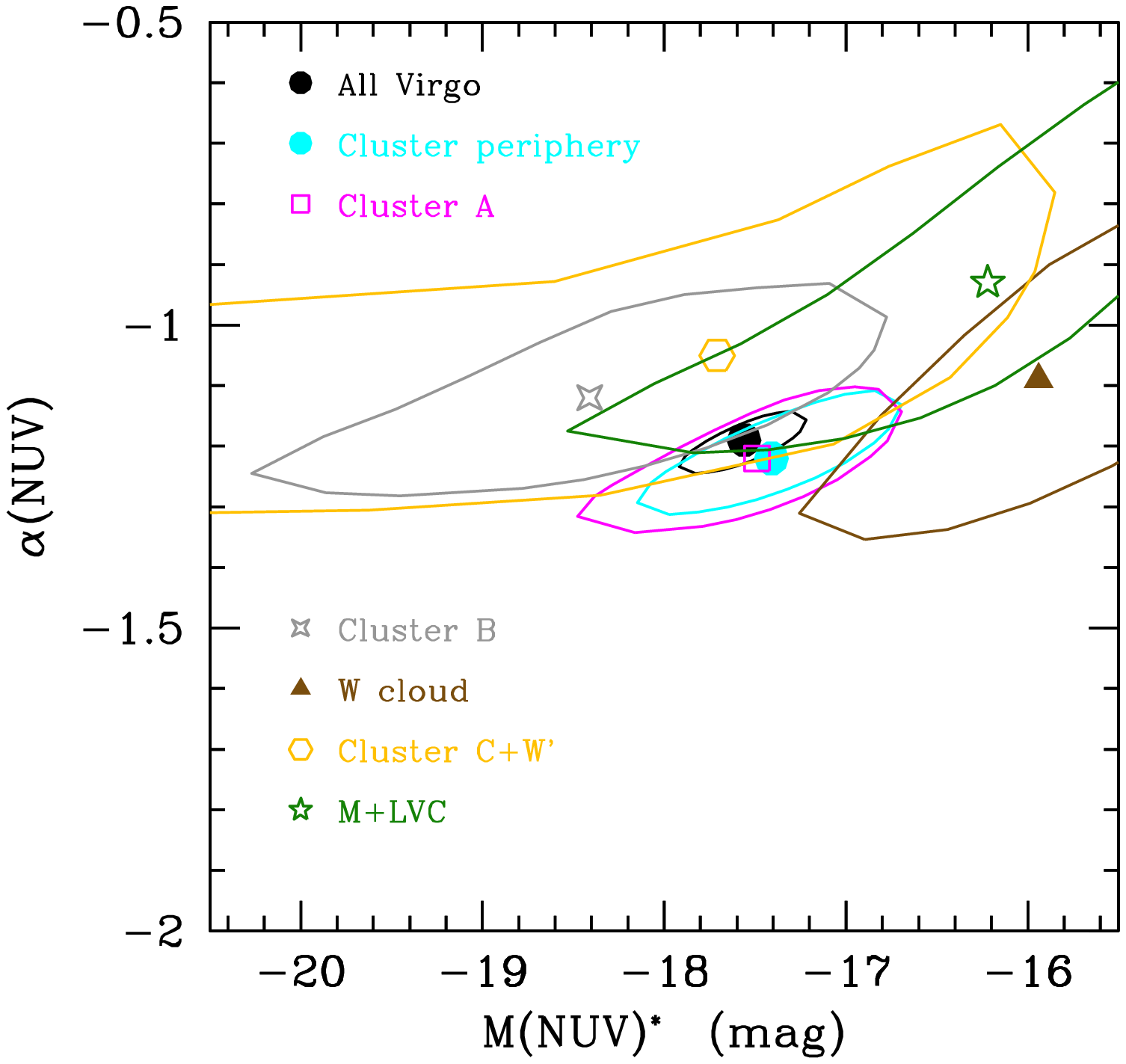}
   \caption{Comparison of the Schechter function parameters determined for the whole Virgo cluster region (black filled circle), for 
   the cluster periphery (cyan filled circle), cluster A (magenta empty square), cluster B (grey empty star), W cloud (brown filled
   triangle), cluster C and W' cloud (orange empty hexagon), and M and LVC clouds (dark green empty star). The contour show
   the 1 sigma probability distribution. }
   \label{LFparareg}%
   \end{figure}

The luminosity function of the main body of the cluster (cluster A) determined within half of the virial radius $R_{vir}$, and that of the cluster periphery ($\gtrsim$ 1.1
$R_{vir}$) can also be compared to those determined for the other cluster substructures. This can be done only for 
cluster B and W cloud, which have a sufficient number of objects. To increase the statistics in the remaining substructures, we combine their data 
according to the general properties of their members as determined in Boselli et al. (2014a). We thus combine cluster C and W' cloud, which are dominated by 
quiescent objects, while the M cloud with the low velocity cloud (LVC) which, on the contrary, are mainly composed of star forming systems.
The comparison of the different NUV luminosity functions is shown in the lower panel of Fig. \ref{LFFTall} and in Fig. \ref{LFparareg}, 
while the fitted parameters are given in Table \ref{LFfit}.
Despite the poor statistics, the comparison of the different luminosity functions suggests that these substructures have a 
flatter slope at the faint end than cluster A or the field. These trends, however, should be confirmed on stronger statistical basis.

\subsection{The luminosity function as a function of morphological type}

The sample can also be cut in different subclasses according to the morphological type of galaxies. This can be done either 
using the accurate morphological classification given in the VCC (Binggeli et al. 1985) or in the EVCC (Kim et al. 2014), or using the colour of galaxies
as done in Boselli et al. (2014a). Consistently with our previous work we select galaxies belonging to the red sequence, the green valley, 
and the blue cloud according to their location in the $M_{star}$ vs. $NUV-i$ diagram using the same selection criteria given in Boselli et al. (2014a).
We prefer to select galaxies according to a quantitative colour-stellar mass relation rather than to a subjective morphology classification
mainly because of the physical interpretation of this diagram in the framework of galaxy transformation in high-density environments 
(see Boselli et al. 2014a for details).

  \begin{figure}
   \centering
   \includegraphics[width=16cm]{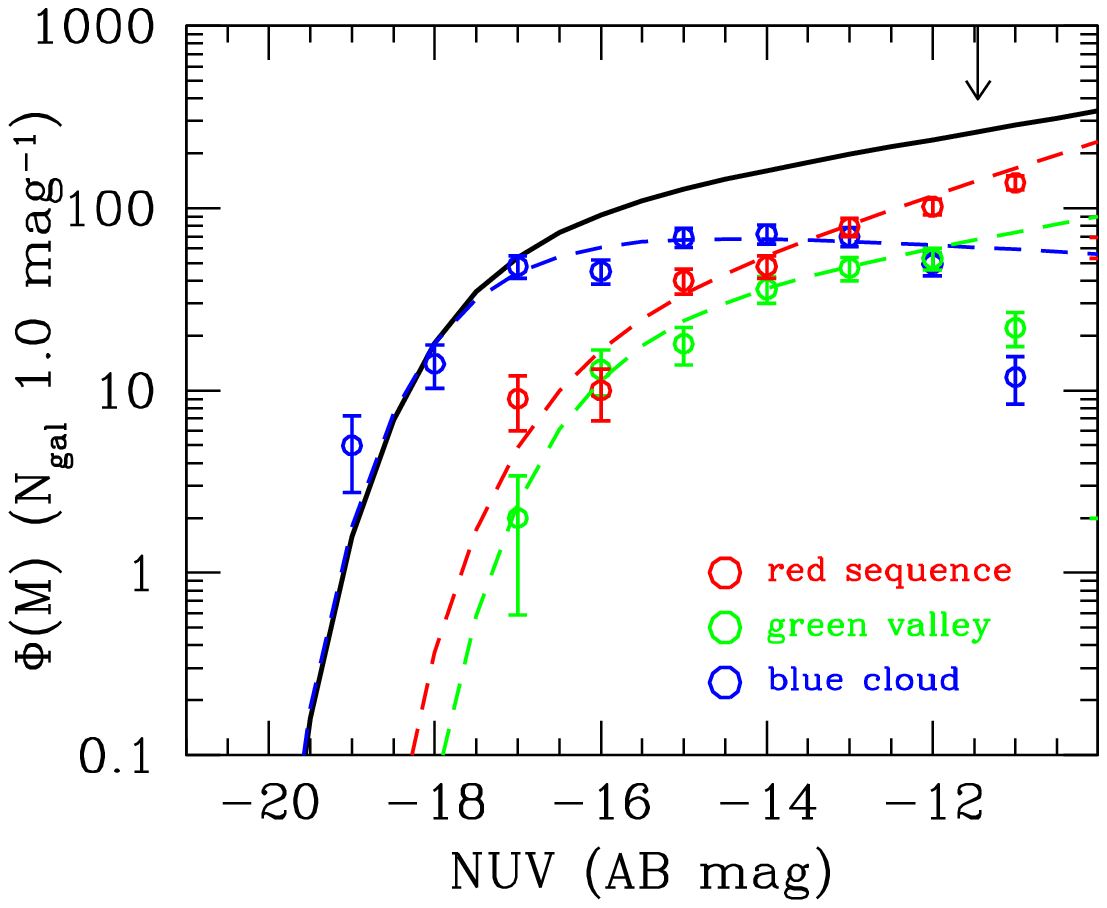}
   \caption{The NUV luminosity function of the Virgo 
   cluster for galaxies of different morphological type selected according to their $NUV-i$ colour, as described in Boselli et al.
   (2014a). Red, green, and blue symbols stand for galaxies in the red sequence, green valley, and blue could, respectively.
   The black solid line shows the total NUV luminosity function of the cluster. The vertical arrow shows the completeness limit of the survey.}
   \label{LFFTtype}%
   \end{figure}

\begin{figure}
   \centering
   \includegraphics[width=10cm]{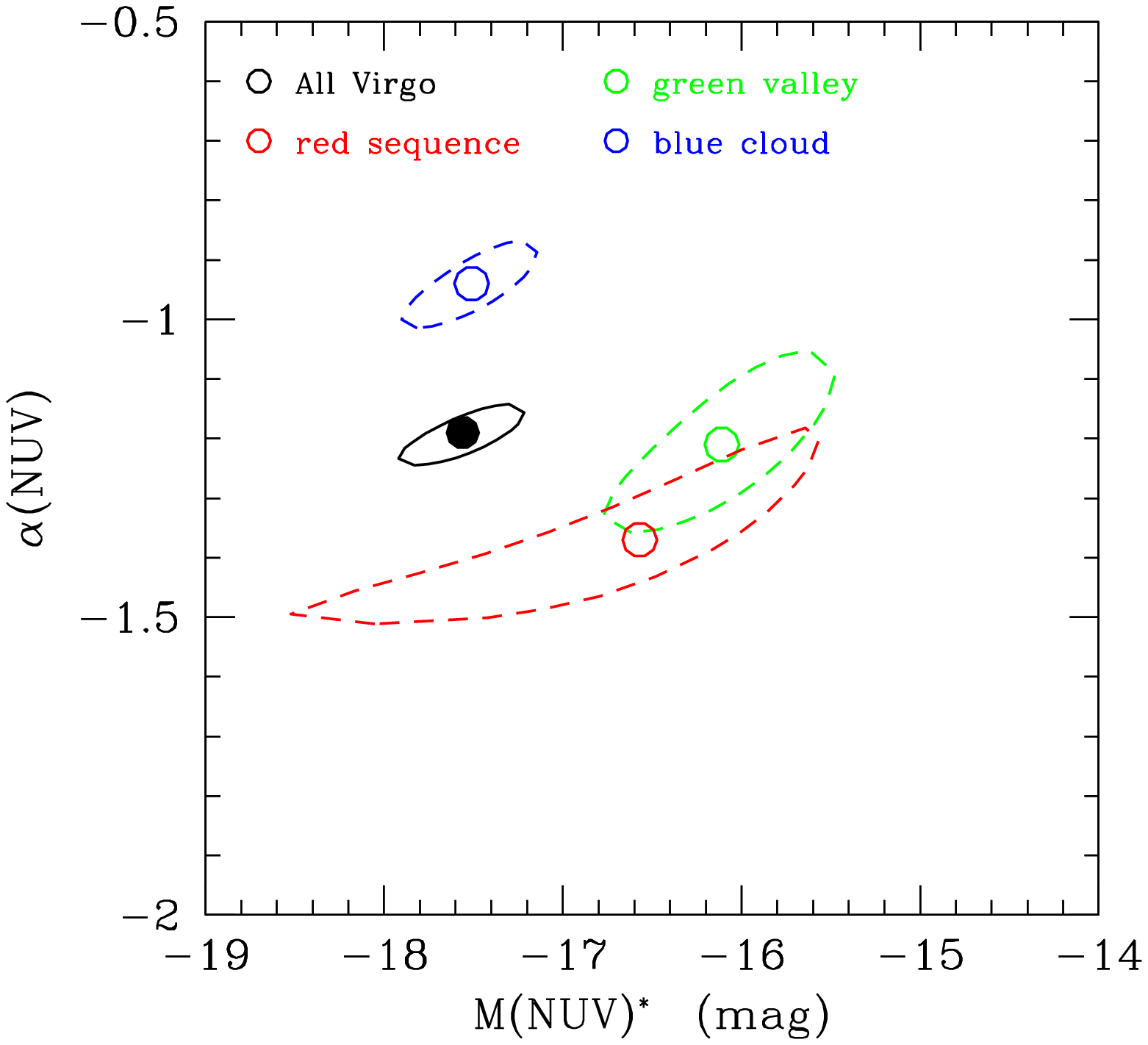}
   \caption{Comparison of the Schechter function parameters determined for the whole Virgo cluster region (black symbol), 
   and for galaxies belonging to the red sequence, green valley, and blue cloud, as defined in Boselli et al. (2014a). The contour show
   the 1 sigma probability distribution. }
   \label{LFparatype}%
   \end{figure}

Figure \ref{LFFTtype} shows the NUV luminosity function of the whole cluster separately for galaxies belonging to the red sequence, the green 
valley, and the blue cloud. The values are given in Table \ref{LFtab}, while the parameters of the best fitted Schechter functions are
given in Table \ref{LFfit} and compared in Fig. \ref{LFparatype}.
Figures \ref{LFFTtype} and \ref{LFparatype}, and Tables \ref{LFtab} and \ref{LFfit}, consistently indicate that at the bright end the NUV luminosity function 
of the Virgo cluster is dominated by late-type galaxies (the difference in $M^*$ between blue and red galaxies is of $\sim$ 1 mag), 
while the increasing number of galaxies at the faint end is mainly due to the red quiescent galaxy population, as already found in other clusters 
(Cortese et al. 2005, 2008). Indeed the slope of the early-type 
red galaxies ($\alpha$ = -1.37) is significantly steeper than the one of the star forming systems within the blue cloud ($\alpha$ = -0.94).

\begin{figure*}
   \centering
   \includegraphics[width=17cm]{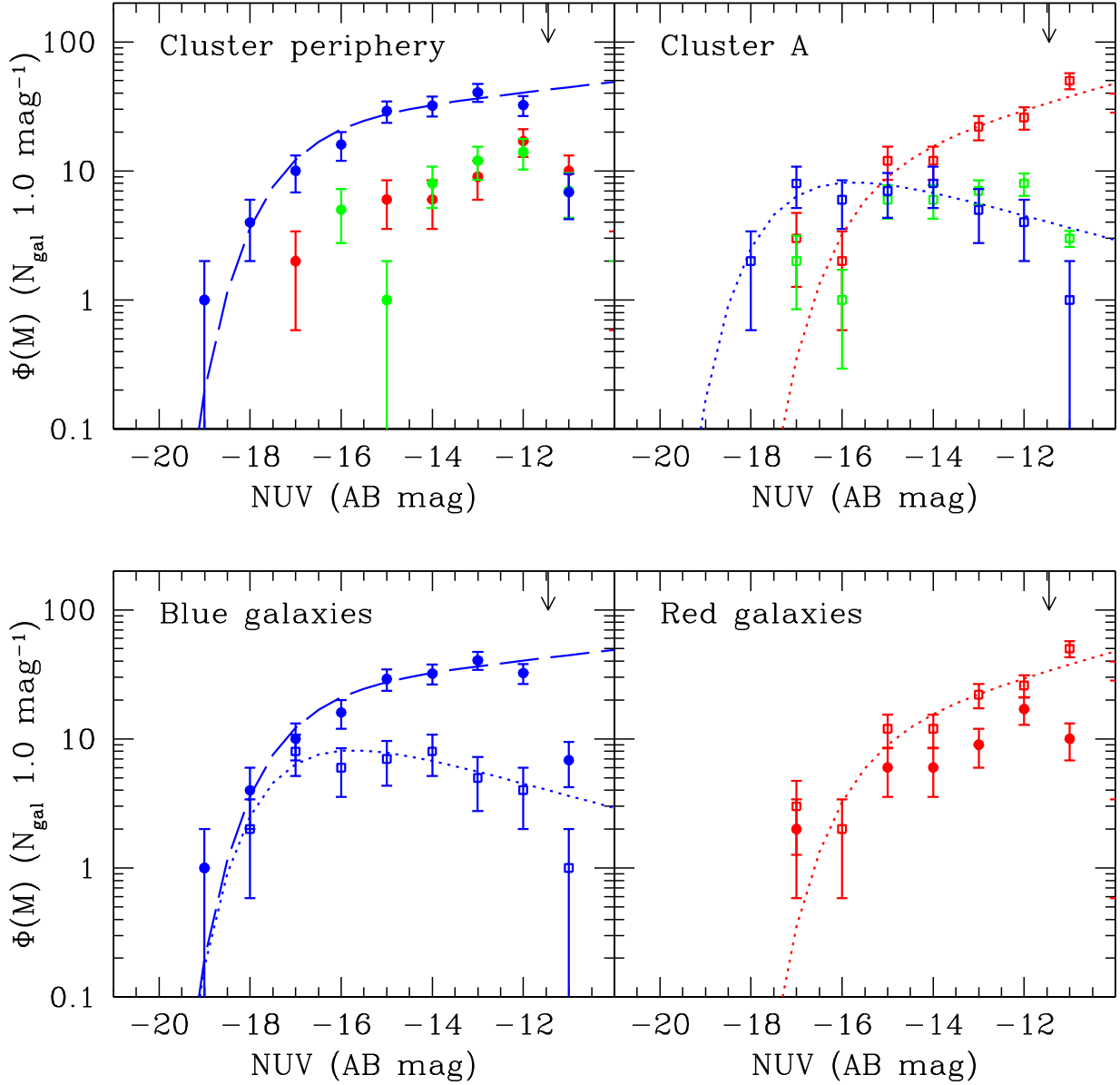}
   \caption{Upper panels: the NUV luminosity function of galaxies in the Virgo cluster periphery (left, filled dots)
   and in the core of cluster A (right, empty squares). Red symbols are for galaxies belonging to the red sequence, 
   green symbols for galaxies in the green valley, and
   blue symbols for galaxies in the blue cloud. Lower panels: the NUV luminosity function of cluster A (empty squares) and cluster periphery (filled dots) galaxies in the blue (left) 
   and red (right) sequences. The red and blue lines give the the best fit Schechter functions derived for the different samples.
   The different luminosity functions have been normalised to $\sim$ the same number of objects. }
   \label{LFcol}%
   \end{figure*}

The comparison between the different morphological classes can also be done separately for galaxies located within cluster A and 
at the periphery of the cluster (Fig. \ref{LFcol}; Table \ref{LFtab}). The limited number of galaxies prevents an accurate determination of 
the parameters of the Schechter function for early-type galaxies in the periphery of the cluster. 
 
Despite a significant decrease of the total number of objects, the shape of the NUV luminosity function of red galaxies does not change significantly between cluster A 
and the periphery. A Kolmogorov-Smirnov test indicates that the probability that the cluster A and the cluster periphery early-type galaxy distributions are driven by the same parent
distribution is 11\%. Important variations are instead observed in the NUV luminosity function of blue galaxies. Evident is the change in the slope at faint luminosities,
from $\alpha$ = -1.10 in the outskirts of the cluster to $\alpha$ = -0.76 in the core of cluster A. In the cluster periphery the NUV luminosity function is dominated by blue
galaxies at all luminosities (notice that the value of $\alpha$ = -1.10 determined for this category of objects is very close to the one derived in the field by Wyder et al.
(2005)), while in the core of the cluster is dominated by the steep rise of red early-type dwarf galaxies below $NUV>$ -14 mag ($\alpha$ = -1.25). This result does
not necessary contradicts what found in the Coma cluster, i.e. that the shape of the NUV luminosity function of early- and late-type galaxies
does not change in the clustercentric distance range 
0.5$^o$ $<$ $R$ $<$ 2$^o$ (Cortese et al. 2008). Indeed, once considered in terms of virial radii, this range (0.3 $<$ $R/R_{Vir}$ $<$ 1.2) 
does not match with the one considered in this analysis (cluster A: $R/R_{Vir}$  $<$ 0.5; cluster periphery: $R/R_{Vir}$  $>$ 1.1).

\subsection{The star formation rate}

The UV emission of star forming systems is due to young and massive stars whose life on the main sequence is relatively short. If the star formation activity
of galaxies is fairly constant over timescales as long as the typical age of the emitting stars (stationarity conditions), UV luminosities corrected for dust attenuation 
can be converted into star formation rates (e.g. Kennicutt 1998, Boselli et al. 2009; Boquien et al. 2014). The constant of transformation can be 
determined using different population synthesis models, and depends on the assumed IMF and metallicity (e.g. Boselli 2011).
The NUV emission of late-type galaxies is principally due to stars with a typical age of $\sim$ 150 Myr (Boissier 2013). These timescales are 
comparable to the timescales for the observed variations in the star formation activity of cluster galaxies.
Indeed, in these objects the interaction with the hostile cluster environment is able to remove, on relatively short timescales 
($\sim$ 100-700 Myr, Boselli et al. 2006, 2008a; Tonnesen \& Bryan 2009), a large fraction of the atomic and molecular gas content (Fumagalli et al. 2009; Boselli et al. 2014b) 
quenching the activity of star formation. This process is particularly efficient 
in dwarf systems, where the shallow potential well of the galaxies cannot keep the gas reservoir anchored to the disc even in the central regions.
The stationarity conditions required for transforming NUV luminosities into star formation rate are thus not necessarily satisfied. 
For comparison with other works and keeping in mind this caveat, however, we can try to convert the NUV luminosities corrected for dust attenuation (see sect. 2) 
into $SFR$. 
Because of the reason mentioned above (quenching of the star formation activity of cluster galaxies), 
using NUV data we expect to overestimate the present day star formation activity of cluster galaxies, in particular in low luminosity objects. 

   \begin{figure}
   \centering
   \includegraphics[width=17cm]{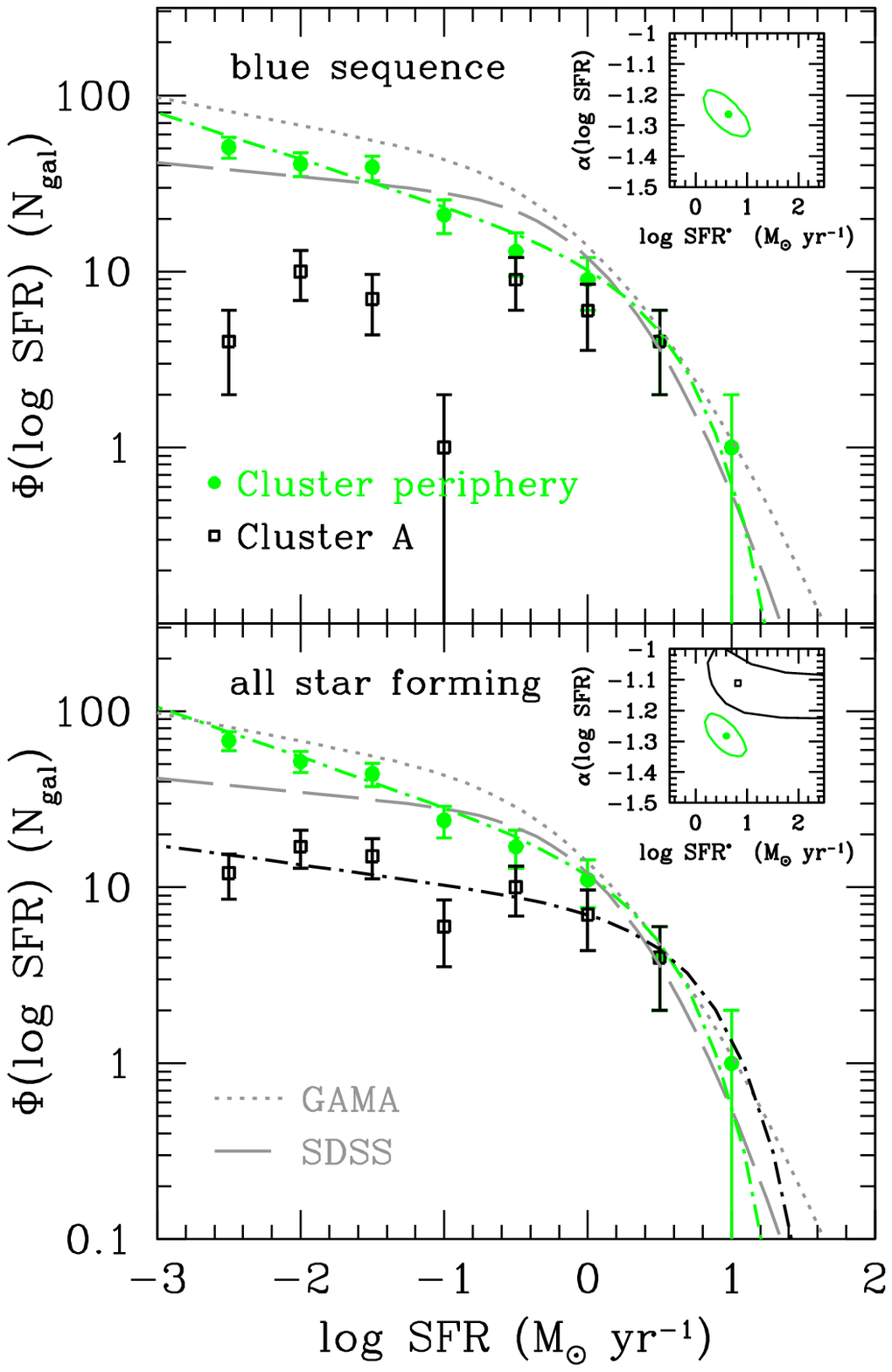}
   \caption{The SFR luminosity function of galaxies in the cluster periphery (green filled dots) and in cluster A 
   (black empty squares) derived from the NUV luminosity function of galaxies in the blue sequence (upper panel) 
   and of all star forming objects, including those in the green valley (lower panel). These SFR luminosity functions 
   (green for the periphery and black for cluster A) dotted-dashed lines
   are compared to those derived by Gunawardhana et al. (2013) for the GAMA (grey dotted line)
   and the SDSS (grey long dashed line) samples. 
   The different luminosity functions have been normalised to $\sim$ the same number of objects for comparison. The small panels in the top right 
   indicate the 1 sigma probability distribution of the fitted Schecheter function parameters.}
   \label{LFSFR}%
   \end{figure}

Figure \ref{LFSFR} shows the distribution of the star formation rate determined for late-type galaxies in the Virgo cluster region. This distribution has
been determined for galaxies in the blue sequence only (upper panel), or for all star forming galaxies (blue sequence and green valley; lower panel) separately for objects situated within cluster A
and at the periphery of the cluster. We do not consider here red quiescent galaxies, where the UV emission can be due to evolved stars (e.g. O'Connel 1999; Boselli et
al. 2005b). 
The SFR luminosity distribution of both the blue sequence and of all the star forming galaxies at the periphery of the Virgo cluster is similar
to the one determined for the field using SDSS and GAMA data (Gunawardhana et al. 2013). It is, however, significantly steeper than the one determined for galaxies located within cluster A,
consistent with what shown in Fig. \ref{LFcol}. These distributions are fitted with a Schechter function whose parameters are given in Table \ref{SFRtabfit}.

\section{Discussion}

The origins of the differences observed with other clusters and with the field (sect. 4.1) might be several, as already discussed in Boselli et al. (2011). They include:\\ 
a) A different sampling in absolute magnitude for the different clusters and the field. The limiting magnitude in Coma, A1367, and Shapley
is only $NUV$ $\leq$ -14.5 vs. $\simeq$ -11.5 in Virgo. If we limit the comparison with these clusters down to the limiting magnitude of $NUV$ $\leq$ -14.5 mag, however, 
we do not observe any significant change in the slope $\alpha$ of the Schechter function in Virgo. Given the degeneracy between
the Schechter parameters $\alpha$ and $M^*$ it is difficult to establish whether the position of the knee of the Schechter function is
significantly different in the various clusters and in the field or is just due to their steeper slope at the faint end. We recall, however, 
that clusters such as Coma and A1367 contain a number of gas-rich late-type systems characterised by a strong activity of star formation probably
induced by their interaction with the surrounding intracluster medium (Iglesias-Paramo et al. 2004). 
These galaxies have been first discovered by Bothun \& Dressler (1986) in the Coma cluster, and are typical of unrelaxed clusters such as A1367. 
Typical examples are the galaxy CGCG 97-73, 97-79, and 97-87 at the north west periphery of the cluster (Gavazzi et al. 2001).
Similar objects are apparently lacking in Virgo as revealed by the complete H$\alpha$ narrow band imaging survey of the cluster
(Koopmann et al. 2001; Boselli \& Gavazzi 2002; Boselli et al. 2002; Gavazzi et al. 2006). \\
b) Boselli et al. (2011) invoked a possible bias in the determination of the UV luminosity function in the central 12 deg.$^2$ of Virgo because of the lack of bright galaxies 
in their sample. The region analysed in this work is $\sim$ 25 times larger and the number of objects has increased by a factor of $\sim$ 8 with respect to that
work, making our determination of the NUV luminosity function robust.\\
c) Possible radial effects in the galaxy distribution. The Coma cluster NUV luminosity function of Cortese et al.
(2008) has been determined over an area of 25 Mpc$^2$ but avoiding the core of the cluster because of the presence of bright stars saturating 
the detector. On the contrary the NUV luminosity function of A1367 is limited to the central $\sim$ 2 Mpc$^2$, corresponding to $\sim$ 0.37 $R_V$.
The properties of the fitted Schechter functions of these two clusters, however, are very similar, as depicted in Figs. \ref{LFFT} and \ref{LFpara}.
If limited to the infalling region in the south west of
Coma, where UV data are available for galaxies down to $NUV$ = -10.5, Hammer et al. (2012) determined a slope $\alpha$ = -1.39, slightly flatter than the
one determined by Cortese et al. (2008) in the outer regions but still significantly steeper than the one determined in this work. Furthermore, we have shown that the 
shape of the fitted Schechter function does not change significantly from the core to the periphery of the cluster, corroborating the idea that radial effects 
are not important.\\
d) The determination of the slope of the UV luminosity function at the faint end strongly depends on poorly constrained statistical corrections 
in clusters other than Virgo (e.g. Rines \& Geller 2008; see Fig \ref{histoNUV}). At the limiting observed magnitude of 21 mag in the NUV used by Cortese et al. (2008)
not all galaxies have an optical counterpart necessary for the star-galaxy identification on the SDSS (Voyer et al. 2014). In the Coma cluster and in the Shapley
supercluster
the completeness in redshift is present only down to $NUV$ = -17, and drops to $\lesssim$ 20\% ~ at the limiting absolute mag of $NUV$ = -14 (Fig. \ref{histoNUV}). 
In Hammer et al. (2012) the UV selected galaxies ($NUV$ $\leq$ 24.5 mag) must have a SDSS optical counterpart with $r$ $\leq$ 21.2 mag. 
With this condition all star forming objects are automatically excluded by the sample.
Although we cannot exclude real differences in the luminosity distribution of galaxies within these clusters, which are known to be characterised by
different intrinsic properties (relaxation state and presence of substructures, spiral fraction, density of the intracluster medium and X-ray properties),
it is conceivable that part of the observed differences results from the uncertain statistical corrections required in clusters more distant than Virgo. 

The NUV luminosity function determined in this work can also be compared to those obtained at other frequencies in the Virgo cluster. As summarised in Boselli \& Gavazzi (2014),
these are now available only in optical bands (Sandage et al. 1985, photographic $B$-band; Trentham \& Hodgkin 2002, $B$-band;  Rines \& Geller 2008, $r$-band; Lieder et al. (2012), 
$V$- and $I$-bands). We can compare our results to those obtained by Rines \& Geller (2008) in the $r$-band within 1 Mpc from M87 since their sampled 
region fairly corresponds to the cluster A definition adopted in this work ($<$ 0.8 Mpc). Furthermore both works use membership criteria based on spectroscopic redshifts
and are thus only marginally affected by statistical corrections.
The fit with a Schechter function done by Rines \& Geller (2008) gives $M_R^*$ = -21.32 and $\alpha$ = -1.28. Considering a typical $NUV-r$ colour of $\sim$ 5.5, 4.5, and 3.5 mag for massive red sequence, green valley,
and blue cloud galaxies as derived from the $NUV-i$ colour-magnitude relation of Boselli et al. (2014a) and the typical $r-i$ colours of nearby galaxies (Fukujita et al. 1995),
we can tentatively transform the NUV luminosity function of cluster A given in Table \ref{LFfit} in a $r$-band luminosity function. We thus expect that the derived $r$-band luminosity function
is dominated by the early-type galaxy population with $M_r^*$ and $\alpha$ values close to those obtained by Rines \& Geller (2008). We recall, however, that 
this is a very crude approximation since i) the colour of galaxies is luminosity-dependent and ii) the $r$- and NUV-bands are sensitive to very different stellar populations
thus selection effects should be important.

The variations of the NUV luminosity function with morphological type and cluster location can be analysed in the framework of galaxy evolution in high-density environments (see Boselli \& Gavazzi 2006, 2014 for a
review). The fact that the 
NUV luminosity function does not change significantly in terms of faint end slope going from the cluster core ($\leq$ 0.5 $R_{vir}$) to the outskirts 
($\gtrsim$ 1.1 $R_{vir}$) and to the field, suggests that whatever perturbing mechanism is acting on low luminosity galaxies entering the cluster, 
this does not change their number per unit NUV luminosity. The interpretation of this observational result, however, is not straightforward since 
the different environmental processes can modify in a complex way both the number of objects and the luminosity of the perturbed galaxies. 
As a first indication we would be tempted to exclude for Virgo the combined effect of dwarf tidal disruption in the core of the cluster and galaxy
harassment invoked by Popesso et al. (2006), Barkhouse et al. (2007), and de Filippis et al. (2011) to explain the observed flattening in the core and the steepening in 
the periphery in the optical luminosity function of nearby clusters. Indeed, we do not observe in Virgo any radial variations of the luminosity function
both in the NUV and in the $r$-band (Rines \& Geller 2008). On the contrary, our results seem consistent with a mild transformation of star forming systems into
quiescent dwarf elliptical galaxies after a rapid ram pressure stripping event as first proposed by Boselli et al. (2008a,b). Because of their shallow potential well
low luminosity star forming galaxies easily lose their gas because of ram pressure stripping and quench their activity of star formation on timescales of $\lesssim$ 1 Gyr. This happens at the periphery of 
the cluster once they encounter the intergalactic medium trapped within the potential of the cluster. Because this timescale is relatively short compared to the 
crossing time of the cluster ($\sim$ 1.7 Gyr), these newly accreted galaxies are already red when they reach the cluster core (Boselli et al. 2014a). This effect is
consistent with the 
observed reversed shape of the NUV luminosity function of red and blue galaxies in the cluster core and at the periphery (Fig.\ref{LFcol}). It is also consistent with 
the spectrophotometric, structural (Boselli et al. 2008a,b), and kinematic (Toloba et al. 2009, 2011, 2012, 2015) properties of Virgo cluster dwarf elliptical galaxies,
as extensively reviewed in Boselli \& Gavazzi (2014).\\

The perturbations induced by the cluster environment on the star formation activity of galaxies are obviously more efficient in low-mass systems than in massive spirals.
In massive objects the steep potential well can retain some gas in the inner regions producing radially truncated discs as in the case of NGC 4569 (Boselli et al. 2006).
The timescale necessary for the total ablation of the atomic and molecular gas here is longer than in dwarfs as indicated both by observations and hydrodynamic simulations (Roediger \&
Bruggen 2007, Tonnesen \& Bryan 2009, Hughes \& Cortese 2009, Cortese \& Hughes 2009, Fumagalli et al. 2009, Gavazzi et al. 2013a,b, Boselli et al. 2014a,b).
The observed increasing difference with decreasing star formation rate in the $SFR$ luminosity function of the Virgo cluster shown in Fig. \ref{LFSFR} is a further 
statistical evidence of this phenomenon.

\section{Conclusion}

We have determined the properties of the NUV luminosity function of the Virgo cluster  over $\sim$ 300 deg.$^2$ 
($\sim$ up to 1.8 virial radii) thanks to the GUViCS survey using a complete sample of $\sim$ 833 galaxies, out of which 808 identified as Virgo members
using spectroscopic data and 10 with optical scaling relations, down to the 100\% ~ completeness limit of $NUV$ = -11.5.
The results of this analysis can be summarised as follows:\\
1) The NUV luminosity function can be represented by a Schechter function with $M^*$ = -17.56 and $\alpha$ = -1.19. $M^*$ is $\sim$ 1-2 mag fainter 
and the slope significantly flatter than the one found in other clusters. $M^*$ is only $\sim$ 0.8 mag fainter and $\alpha$ 
comparable to those determined in the field. The difference with other clusters is probably due to the quite uncertain statistical correction
necessary to identify members in clusters at distances higher than Virgo.\\
2) The NUV luminosity function of galaxies of different morphological type as determined according to their position within the $NUV-i$ vs. $M_{star}$ 
relation is significantly different. Late-type galaxies belonging to the blue sequence dominate the bright end of the luminosity function,
while their faint end is very flat $\alpha$ = -0.94). On the contrary, early-type galaxies belonging to the red sequence dominate
the faint end ($\alpha$ = -1.37) of the total NUV luminosity function below $NUV$ $\simeq$ -13 mag.\\
3) The overall properties of the NUV luminosity function determined within 0.5 virial radii of cluster A, the main structure of Virgo, and in 
the cluster periphery, outside $\sim$ 1.1 virial radii, are comparable. In the outskirts of the cluster the NUV luminosity function is 
comparable to the one determined in the field by Wyder et al. (2005). Here it is dominated down to $NUV$ $\simeq$ -11.5 by blue galaxies, with a mild rise at 
faint luminosities ($\alpha$=-1.10). In the cluster core red quiescent galaxies dominate for $NUV$ $\gtrsim$ -15 mag ($\alpha$ = -1.25), while the number of
star forming systems significantly drops ($\alpha$ = -0.76).
These observed properties of the NUV luminosity function are consistent with a rapid transformation of star forming galaxies recently entered in the cluster
environment after a ram-pressure stripping event.

\begin{acknowledgements}

We thank the anonymous referee for constructive comments that helped improving the quality of the manuscript.
This research has been financed by the French ANR grant VIRAGE and the French national program PNCG.
We wish to thank the GALEX Time Allocation Committee for the generous allocation of time devoted to this project.
LC acknowledges financial support from the Australian Research Council (DP130100664).
This research has made use of the NASA/IPAC Extragalactic Database (NED) 
which is operated by the Jet Propulsion Laboratory, California Institute of 
Technology, under contract with the National Aeronautics and Space Administration
and of the GOLDMine database (http://goldmine.mib.infn.it/) (Gavazzi et al. 2003).
GALEX (Galaxy Evolution Explorer) is a NASA Small Explorer, launched in 2003 April. We gratefully acknowledge NASA support for construction, operation, and 
science analysis for the GALEX mission, developed in cooperation with the Centre National d'Etudes Spatiales of France and the Korean Ministry of Science and Technology.
Funding for the SDSS and SDSS-II has been provided by the Alfred P. Sloan Foundation, the Participating Institutions, the National
Science Foundation, the U.S. Department of Energy, the National Aeronautics and Space Administration, the Japanese Monbukagakusho, the
Max Planck Society, and the Higher Education Funding Council for England. The SDSS Web Site is http://www.sdss.org/.
The SDSS is managed by the Astrophysical Research Consortium for the Participating Institutions. The Participating Institutions are the
American Museum of Natural History, Astrophysical Institute Potsdam, University of Basel, University of Cambridge, Case Western Reserve
University, University of Chicago, Drexel University, Fermilab, the Institute for Advanced Study, the Japan Participation Group, Johns
Hopkins University, the Joint Institute for Nuclear Astrophysics, the Kavli Institute for Particle Astrophysics and Cosmology, the Korean
Scientist Group, the Chinese Academy of Sciences (LAMOST), Los Alamos National Laboratory, the Max-Planck-Institute for Astronomy (MPIA), the
Max-Planck-Institute for Astrophysics (MPA), New Mexico State University, Ohio State University, University of Pittsburgh,
University of Portsmouth, Princeton University, the United States Naval Observatory, and the University of Washington.

\end{acknowledgements}

\begin{table*}
\caption{The NUV luminosity function. }
\label{LFtab}
{
\[
\begin{tabular}{ccccc}
\hline
\noalign{\smallskip}
\hline
      NUV     &  All      & Blue       & Green     & Red   \\
\hline
         -20  &      -    &      -     &       -   &	   -	 \\
         -19  &      5    &      5     &       -   &	   -	 \\
         -18  &     14    &     14     &       -   &	   -	 \\
         -17  &     59    &     48     &       2   &	   9	 \\
         -16  &     68    &     45     &      13   &	  10	 \\
         -15  &    127    &     69     &      18   &	  40	 \\
         -14  &    156    &     72     &      36   &	  48	 \\
         -13  &    196    &     70     &      47   &	  79	 \\
         -12  &    204    &     49     &      53   &	 102	 \\
\noalign{\smallskip}
\hline
\end{tabular}
\]
Note: number of objects
}
\end{table*}

\begin{table*}
\caption{The NUV luminosity function in the different cluster substructures. }
\label{LFcloudstab}
{
\[
\begin{tabular}{ccccccc}
\hline
\noalign{\smallskip}
\hline
      NUV     & Cluster A & Cluster B & W cloud & C + W' & M + LVC   & Periphery \\
\hline
         -20  &      -    &	  -   &      -  &     -  &	  -  &     -  \\
         -19  &      -    &	  1   &      -  &     -  &	  -  &     1  \\
         -18  &      2    &	  -   &      -  &     -  &	  -  &     4  \\
         -17  &     13    &	  5   &      1  &     3  &	  2  &    12  \\
         -16  &      9    &	 10   &      -  &     4  &	  4  &    21  \\
         -15  &     25    &	  6   &      7  &     7  &	  6  &    36  \\
         -14  &     26    &	 10   &     12  &     4  &	  9  &    46  \\
         -13  &     34    &	 13   &     19  &     8  &	 14  &    62  \\
         -12  &     38    &	 11   &     14  &     6  &	  6  &    63  \\
\noalign{\smallskip}
\hline
\end{tabular}
\]
Note: number of objects
}
\end{table*}

\begin{table*}
\caption{The NUV luminosity function: cluster A vs. periphery }
\label{LFtypetab}
{
\[
\begin{tabular}{ccccccc}
\hline
\noalign{\smallskip}
\hline
      NUV     &Cluster A red & Cluster A green & Cluster A blue & Periphery red & Periphery green & Periphery blue \\
\hline
         -20  &          -   &       -   &      -   &	    -  &        -   &	    - \\
         -19  &          -   &       -   &      -   &	    -  &        -   &	    1 \\
         -18  &          -   &       -   &      2   &	    -  &        -   &	    4 \\
         -17  &          3   &       2   &      8   &	    2  &        -   &	   10 \\
         -16  &          2   &       1   &      6   &	    -  &        5   &	   16 \\
         -15  &         12   &       6   &      7   &	    6  &        1   &	   29 \\
         -14  &         12   &       6   &      8   &	    6  &        8   &	   32 \\
         -13  &         22   &       7   &      5   &	    9  &       12   &	   41 \\
         -12  &         26   &       8   &      4   &	   17  &       14   &	   32 \\
\noalign{\smallskip}
\hline
\end{tabular}
\]
Note: number of objects
}
\end{table*}

\bgroup
\begin{table*}
\caption{The best fitting parameters for the NUV luminosity function. } 
\label{LFfit} 
{\tiny
\[
\begin{tabular}{lccc}
\hline
\noalign{\smallskip}
\def\arraystretch{2.7} %
Sample & $\alpha$  & $M^*$ & $\phi_*$ \\
\hline
\vspace{1 mm}
Full sample & $-1.19_{-0.03}^{+0.03}$ & $-17.56_{-0.24}^{+0.23}$ & $88.40_{-14.71}^{+16.33}$\\ %
\vspace{1 mm}
Red sequence & $-1.37_{-0.10}^{+0.12}$ & $-16.57_{-0.98}^{+0.70}$ & $24.93_{-13.35}^{+20.59}$\\ %
\vspace{1 mm}
Green valley & $-1.21_{-0.10}^{+0.10}$ & $-16.11_{-0.42}^{+0.40}$ & $27.74_{-9.42}^{+11.88}$\\ %
\vspace{1 mm}
Blue cloud & $-0.94_{-0.05}^{+0.05}$ & $-17.51_{-0.25}^{+0.25}$ & $85.24_{-14.11}^{+15.40}$\\ %
\vspace{1 mm}
Late-type & $-1.07_{-0.04}^{+0.04}$ & $-17.54_{-0.25}^{+0.24}$ & $85.58_{-14.42}^{+15.93}$\\ %
\vspace{1 mm}
Cluster A & $-1.22_{-0.08}^{+0.08}$ & $-17.49_{-0.58}^{+0.50}$ & $14.14_{-5.19}^{+6.37}$\\ %
\vspace{1 mm}
Cluster periphery & $-1.22_{-0.07}^{+0.07}$ & $-17.41_{-0.48}^{+0.47}$ & $25.51_{-8.03}^{+10.45}$\\ %
\vspace{1 mm}
Cluster A red sequence & $-1.25_{-0.28}^{+0.24}$ & $-15.70_{-2.79}^{+0.73}$ & $12.92_{-12.92}^{+13.27}$\\ %
\vspace{1 mm}
Cluster A green valley & $-0.92_{-0.51}^{+0.39}$ & $-15.23_{-0.91}^{+0.72}$ & $10.81_{-9.56}^{+9.62}$\\ %
\vspace{1 mm}
Cluster A blue could & $-0.76_{-0.15}^{+0.16}$ & $-17.29_{-0.64}^{+0.50}$ & $14.62_{-5.45}^{+6.20}$\\ %
\vspace{1 mm}
Cluster A late-type & $-0.99_{-0.11}^{+0.11}$ & $-17.41_{-0.62}^{+0.50}$ & $14.13_{-5.26}^{+6.11}$\\ %
\vspace{1 mm}
Cluster periphery red sequence & $-1.45_{-0.16}^{+0.21}$ & $-23.97_{-14.82}^{+7.25}$ & $0.12_{-0.12}^{+4.46}$\\ %
\vspace{1 mm}
Cluster periphery green valley & $-0.71_{-0.51}^{+0.55}$ & $-13.64_{-0.79}^{+0.58}$ & $28.82_{-17.29}^{+15.14}$\\ %
\vspace{1 mm}
Cluster periphery blue cloud & $-1.10_{-0.08}^{+0.09}$ & $-17.32_{-0.55}^{+0.56}$ & $24.99_{-8.22}^{+11.06}$\\ %
\vspace{1 mm}
Cluster periphery late-type & $-1.15_{-0.07}^{+0.08}$ & $-17.27_{-0.53}^{+0.50}$ & $27.13_{-8.83}^{+11.51}$\\ %
\vspace{1 mm}
Cluster B & $-1.12_{-0.11}^{+0.11}$ & $-18.41_{-0.92}^{+0.94}$ & $6.54_{-3.11}^{+4.05}$\\ %
\vspace{1 mm}
W cloud & $-1.09_{-0.17}^{+0.72}$ & $-15.94_{-0.77}^{+2.13}$ & $12.77_{-6.37}^{+36.81}$\\ %
\vspace{1 mm}
Cluster C + W' cloud & $-1.05_{-0.19}^{+0.24}$ & $-17.71_{-2.36}^{+1.38}$ & $5.28_{-4.91}^{+6.10}$\\ %
\vspace{1 mm}
M cloud + LVC & $-0.93_{-0.19}^{+0.24}$ & $-16.22_{-1.05}^{+0.97}$ & $11.61_{-5.91}^{+8.95}$\\ %
\vspace{1 mm}
NGVS footprint$^a$ & $-1.18_{-0.04}^{+0.04}$ & $-17.50_{-0.27}^{+0.26}$ & $58.78_{-11.66}^{+13.01}$\\ %
\vspace{1 mm}
NGVS footprint$^a$ red sequence & $-1.36_{-0.12}^{+0.15}$ & $-16.31_{-1.07}^{+0.77}$ & $21.57_{-12.58}^{+20.91}$\\ %
\vspace{1 mm}
NGVS footprint$^a$ late-type & $-0.98_{-0.06}^{+0.06}$ & $-17.36_{-0.28}^{+0.26}$ & $64.92_{-12.50}^{+13.60}$\\ %
\noalign{\smallskip}
\hline
\end{tabular}
\]
Note: $a$ = 1 virial radius from M87 and M49.
}
\end{table*}
\egroup

\begin{table*}
\caption{The SFR luminosity function.}
\label{LFSFRtab}
{
\[
\begin{tabular}{ccccc}
\hline
\noalign{\smallskip}
\hline
log SFR         &  Cluster A blue+green& Periphery blue+green & Cluster A blue & Periphery blue \\
\hline
        -2.5   &       12     &     68   &    4 &  51	   \\
        -2.0   &       17     &     52   &   10 &  41	   \\
        -1.5   &       15     &     44   &    7 &  39	   \\
        -1.0   &        6     &     24   &    1 &  21	   \\
   	-0.5   &       10     &     17   &    9 &  13	   \\
         0     &   	7     &     11   &    6 &   9	   \\
         0.5   &   	4     &      4   &    4 &   4	   \\
         1.0   &   	0     &      1   &    0 &   1	   \\
\noalign{\smallskip}
\hline
\end{tabular}
\]
Note: number of objects
}
\end{table*}

\bgroup
\begin{table*}
\caption{The best fitting parameters for the SFR luminosity function. } 
\label{SFRtabfit} 
{\tiny
\[
\begin{tabular}{lccc}
\hline
\noalign{\smallskip}
\def\arraystretch{2.7} %
Sample & $\alpha$  & log$SFR^*$ & $\phi_*$ \\
\hline
\vspace{1 mm}
Cluster periphery, blue cloud & $-1.26_{-0.05}^{+0.05}$ & $0.64_{-0.31}^{+0.27}$ & $8.82_{-3.09}^{+4.66}$\\
\vspace{1 mm}
Cluster periphery, all star forming & $-1.28_{-0.04}^{+0.05}$ & $0.60_{-0.27}^{+0.25}$ & $10.26_{-3.47}^{+4.96}$\\
\vspace{1 mm}
Cluster A, all star forming & $-1.11_{-0.08}^{+0.08}$ & $0.83_{-0.43}^{+0.64}$ & $6.58_{-5.07}^{+4.09}$\\
\noalign{\smallskip}
\hline
\end{tabular}
\]
}
\end{table*}
\egroup


\begin{thebibliography}{}

\bibitem[Abazajian et al.(2009)]{2009ApJS..182..543A} Abazajian, K.~N., Adelman-McCarthy, J.~K., Ag{\"u}eros, M.~A., et al.\ 2009, \apjs, 182, 543 
\bibitem[Barkhouse et al.(2007)]{2007ApJ...671.1471B} Barkhouse, W.~A., Yee, H.~K.~C., \& L{\'o}pez-Cruz, O.\ 2007, \apj, 671, 1471 
\bibitem[Binggeli et al.(1985)]{1985AJ.....90.1681B} Binggeli, B., Sandage, A., \& Tammann, G.~A.\ 1985, \aj, 90, 1681 
\bibitem[Binggeli et al.(1988)]{1988ARA&A..26..509B} Binggeli, B., Sandage, A., \& Tammann, G.~A.\ 1988, \araa, 26, 509 
\bibitem[Boissier(2013)]{2013pss6.book..141B} Boissier, S.\ 2013, Planets, Stars and Stellar Systems.~Volume 6: Extragalactic Astronomy and Cosmology, 141 
\bibitem[Boissier et al.(2012)]{2012A&A...545A.142B} Boissier, S., Boselli, A., Duc, P.-A., et al.\ 2012, \aap, 545, A142 
\bibitem[Boissier et al.(2015)]{} Boissier, S., Boselli, A., Voyer, E., et al., 2015, arXiv:1504.06111
\bibitem[Boquien et al.(2014)]{2014A&A...571A..72B} Boquien, M., Buat, V., \& Perret, V.\ 2014, \aap, 571, AA72 
\bibitem[Boselli(2011)]{2011pvg..book.....B} Boselli, A.\ 2011, A Panchromatic View of Galaxies, by Alessandro Boselli.~- Practical Approach Book - ISBN-10: 3-527-40991-2.~ISBN-13: 978-3-527-40991-4 - Wiley-VCH, Berlin 2011.~XVI, 324pp, Hardcover,  
\bibitem[Boselli \& Gavazzi(2002)]{2002A&A...386..124B} Boselli, A., \& Gavazzi, G.\ 2002, \aap, 386, 124 
\bibitem{2006PASP..118..517B} Boselli, A., \& Gavazzi, G.\ 2006, \pasp, 118, 517 
\bibitem[Boselli \& Gavazzi(2014)]{2014A&ARv..22...74B} Boselli, A., \& Gavazzi, G.\ 2014, \aapr, 22, 74 
\bibitem[Boselli et al.(2002)]{2002A&A...386..134B} Boselli, A., Iglesias-P{\'a}ramo, J., V{\'{\i}}lchez, J.~M., \& Gavazzi, G.\ 2002, \aap, 386, 134 
\bibitem[Boselli et al.(2005)]{2005ApJ...629L..29B} Boselli, A., Cortese, L., Deharveng, J.~M., et al.\ 2005a, \apjl, 629, L29 
\bibitem[Boselli et al.(2005)]{2005ApJ...623L..13B} Boselli, A., Boissier, S., Cortese, L., et al.\ 2005b, \apjl, 623, L13 
\bibitem{2006ApJ...651..811B} Boselli, A., Boissier, S., Cortese, L., et al.\ 2006, \apj, 651, 811 
\bibitem{2008ApJ...674..742B} Boselli, A., Boissier, S., Cortese, L., \& Gavazzi, G.\ 2008a, \apj, 674, 742 
\bibitem{2008A&A...489.1015B} Boselli, A., Boissier, S., Cortese, L., \& Gavazzi, G.\ 2008b, \aap, 489, 1015 
\bibitem[Boselli et al.(2009)]{2009ApJ...706.1527B} Boselli, A., Boissier, S., Cortese, L., et al.\ 2009, \apj, 706, 1527 
\bibitem{2011A&A...528A.107B} Boselli, A., Boissier, S., Heinis, S., et al.\ 2011, \aap, 528, A107 
\bibitem[Boselli et al.(2014)]{2014A&A...570A..69B} Boselli, A., Voyer, E., Boissier, S., et al.\ 2014a, \aap, 570, AA69 
\bibitem[Boselli et al.(2014)]{2014A&A...564A..67B} Boselli, A., Cortese, L., Boquien, M., et al.\ 2014b, \aap, 564, A67 
\bibitem[Bothun \& Dressler(1986)]{1986ApJ...301...57B} Bothun, G.~D., \& Dressler, A.\ 1986, \apj, 301, 57 
\bibitem[Byrd \& Valtonen(1990)]{1990ApJ...350...89B} Byrd, G., \& Valtonen, M.\ 1990, \apj, 350, 89 
\bibitem[Cortese \& Hughes(2009)]{2009MNRAS.400.1225C} Cortese, L., \& Hughes, T.~M.\ 2009, \mnras, 400, 1225 
\bibitem[Cortese et al.(2005)]{2005ApJ...623L..17C} Cortese, L., Boselli, A., Gavazzi, G., et al.\ 2005, \apjl, 623, L17 
\bibitem[Cortese et al.(2008)]{2008MNRAS.390.1282C} Cortese, L., Gavazzi, G., \& Boselli, A.\ 2008, \mnras, 390, 1282 
\bibitem[Cowie \& Songaila(1977)]{1977Natur.266..501C} Cowie, L.~L., \& Songaila, A.\ 1977, \nat, 266, 501 
\bibitem[D'Abrusco et al.(2007)]{2007ApJ...663..752D} D'Abrusco, R., Staiano, A., Longo, G., et al.\ 2007, \apj, 663, 752 
\bibitem[Davies et al.(2010)]{2010A&A...518L..48D} Davies, J.~I., Baes, M., Bendo, G.~J., et al.\ 2010, \aap, 518, L48 
\bibitem[de Filippis et al.(2011)]{2011MNRAS.414.2771D} de Filippis, E., Paolillo, M., Longo, G., et al.\ 2011, \mnras, 414, 2771 
\bibitem[Dressler(1980)]{1980ApJ...236..351D} Dressler, A.\ 1980, \apj, 236, 351 
\bibitem[Dressler(2004)]{2004cgpc.symp..206D} Dressler, A.\ 2004, Clusters of Galaxies: Probes of Cosmological Structure and Galaxy Evolution, 206 
\bibitem[Dressler et al.(1997)]{1997ApJ...490..577D} Dressler, A., Oemler, A., Jr., Couch, W.~J., et al.\ 1997, \apj, 490, 577 
\bibitem[Drinkwater et al.(2000)]{2000PASA...17..227D} Drinkwater, M.~J., Jones, J.~B., Gregg, M.~D., \& Phillipps, S.\ 2000, \pasa, 17, 227 
\bibitem[Ferrarese et al.(2012)]{2012ApJS..200....4F} Ferrarese, L., C{\^o}t{\'e}, P., Cuillandre, J.-C., et al.\ 2012, \apjs, 200, 4 
\bibitem[Ferrarese et al.(2015)]{} Ferrarese, L., C{\^o}t{\'e}, P., MacArthur, L., et al., 2015, submitted to \apjs
\bibitem[Fitzpatrick \& Massa(2007)]{2007ApJ...663..320F} Fitzpatrick, E.~L., \& Massa, D.\ 2007, \apj, 663, 320 
\bibitem[Fukugita et al.(1995)]{1995PASP..107..945F} Fukugita, M., Shimasaku, K., \& Ichikawa, T.\ 1995, \pasp, 107, 945 
\bibitem[Fumagalli et al.(2009)]{2009ApJ...697.1811F} Fumagalli, M., Krumholz, M.~R., Prochaska, J.~X., Gavazzi, G., \& Boselli, A.\ 2009, \apj, 697, 1811 
\bibitem[Gavazzi et al.(1999)]{1999MNRAS.304..595G} Gavazzi, G., Boselli, A., Scodeggio, M., Pierini, D., \& Belsole, E.\ 1999, \mnras, 304, 595 
\bibitem[Gavazzi et al.(2001)]{2001ApJ...563L..23G} Gavazzi, G., Boselli, A., Mayer, L., et al.\ 2001, \apjl, 563, L23 
\bibitem[Gavazzi et al.(2003)]{2003A&A...400..451G} Gavazzi, G., Boselli, A., Donati, A., Franzetti, P., \& Scodeggio, M.\ 2003, \aap, 400, 451 
\bibitem[Gavazzi et al.(2006)]{2006A&A...446..839G} Gavazzi, G., Boselli, A., Cortese, L., et al.\ 2006, \aap, 446, 839 
\bibitem[Gavazzi et al.(2013)]{2013A&A...553A..89G} Gavazzi, G., Fumagalli, M., Fossati, M., et al.\ 2013a, \aap, 553, A89 
\bibitem[Gavazzi et al.(2013)]{2013A&A...553A..90G} Gavazzi, G., Savorgnan, G., Fossati, M., et al.\ 2013b, \aap, 553, A90 
\bibitem[Giovanelli et al.(2005)]{2005AJ....130.2598G} Giovanelli, R., Haynes, M.~P., Kent, B.~R., et al.\ 2005, \aj, 130, 2598 
\bibitem[Gunawardhana et al.(2013)]{2013MNRAS.433.2764G} Gunawardhana, M.~L.~P., Hopkins, A.~M., Bland-Hawthorn, J., et al.\ 2013, \mnras, 433, 2764 
\bibitem[Gunn \& Gott(1972)]{1972ApJ...176....1G} Gunn, J.~E., \& Gott, J.~R., III 1972, \apj, 176, 1 
\bibitem[Haines et al.(2011)]{2011MNRAS.412..127H} Haines, C.~P., Busarello, G., Merluzzi, P., et al.\ 2011, \mnras, 412, 127 
\bibitem[Hammer et al.(2012)]{2012ApJ...745..177H} Hammer, D.~M., Hornschemeier, A.~E., Salim, S., et al.\ 2012, \apj, 745, 177 
\bibitem[Hao et al.(2011)]{2011ApJ...741..124H} Hao, C.-N., Kennicutt, R.~C., Johnson, B.~D., et al.\ 2011, \apj, 741, 124 
\bibitem[Hilker et al.(1999)]{1999A&AS..134...75H} Hilker, M., Infante, L., Vieira, G., Kissler-Patig, M., \& Richtler, T.\ 1999, \aaps, 134, 75 
\bibitem[Hughes \& Cortese(2009)]{2009MNRAS.396L..41H} Hughes, T.~M., \& Cortese, L.\ 2009, \mnras, 396, L41 
\bibitem[Kennicutt(1998)]{1998ARA&A..36..189K} Kennicutt, R.~C., Jr.\ 1998, \araa, 36, 189 
\bibitem[Kim et al.(2014)]{2014ApJS..215...22K} Kim, S., Rey, S.-C., Jerjen, H., et al.\ 2014, \apjs, 215, 22 
\bibitem[Koda et al.(2015)]{2015ApJ...807L...2K} Koda, J., Yagi, M., Yamanoi, H., \& Komiyama, Y.\ 2015, \apjl, 807, L2 
\bibitem[Koopmann et al.(2001)]{2001ApJS..135..125K} Koopmann, R.~A., Kenney, J.~D.~P., \& Young, J.\ 2001, \apjs, 135, 125 
\bibitem[Iglesias-P{\'a}ramo et al.(2004)]{2004A&A...421..887I} Iglesias-P{\'a}ramo, J., Boselli, A., Gavazzi, G., \& Zaccardo, A.\ 2004, \aap, 421, 887 
\bibitem[Ilbert et al.(2005)]{2005A&A...439..863I} Ilbert, O., Tresse, L., Zucca, E., et al.\ 2005, \aap, 439, 863 
\bibitem[James \& Roos 1995]{} James, F., Roos, M., 1995, MINUIT Function Minimization and Error Analysis, Version 95.03, CERN Program Library D506
\bibitem[Larson et al.(1980)]{1980ApJ...237..692L} Larson, R.~B., Tinsley, B.~M., \& Caldwell, C.~N.\ 1980, \apj, 237, 692 
\bibitem[Lieder et al.(2012)]{2012A&A...538A..69L} Lieder, S., Lisker, T., Hilker, M., Misgeld, I., \& Durrell, P.\ 2012, \aap, 538, A69 
\bibitem[McLaughlin(1999)]{1999ApJ...512L...9M} McLaughlin, D.~E.\ 1999, \apjl, 512, L9 
\bibitem[Mei et al.(2007)]{2007ApJ...655..144M} Mei, S., Blakeslee, J.~P., C{\^o}t{\'e}, P., et al.\ 2007, \apj, 655, 144 
\bibitem[Merritt(1983)]{1983ApJ...264...24M} Merritt, D.\ 1983, \apj, 264, 24 
\bibitem[Mihos et al.(2015)]{2015ApJ...809L..21M} Mihos, J.~C., Durrell, P.~R., Ferrarese, L., et al.\ 2015, \apjl, 809, L21 
\bibitem[Moore et al.(1998)]{1998ApJ...495..139M} Moore, B., Lake, G., \& Katz, N.\ 1998, \apj, 495, 139 
\bibitem[Morrissey et al.(2007)]{2007ApJS..173..682M} Morrissey, P., Conrow, T., Barlow, T.~A., et al.\ 2007, \apjs, 173, 682 
\bibitem[Nulsen(1982)]{1982MNRAS.198.1007N} Nulsen, P.~E.~J.\ 1982, \mnras, 198, 1007 
\bibitem[O'Connell(1999)]{1999ARA&A..37..603O} O'Connell, R.~W.\ 1999, \araa, 37, 603 
\bibitem[Popesso et al.(2006)]{2006A&A...445...29P} Popesso, P., Biviano, A., B{\"o}hringer, H., \& Romaniello, M.\ 2006, \aap, 445, 29 
\bibitem[Raichoor et al.(2014)]{2014ApJ...797..102R} Raichoor, A., Mei, S., Erben, T., et al.\ 2014, \apj, 797, 102 
\bibitem[Rines \& Geller(2008)]{2008AJ....135.1837R} Rines, K., \& Geller, M.~J.\ 2008, \aj, 135, 1837 
\bibitem[Roediger \& Br{\"u}ggen(2007)]{2007MNRAS.380.1399R} Roediger, E., \& Br{\"u}ggen, M.\ 2007, \mnras, 380, 1399 
\bibitem[Sandage et al.(1985)]{1985AJ.....90.1759S} Sandage, A., Binggeli, B., \& Tammann, G.~A.\ 1985, \aj, 90, 1759 
\bibitem[Schlegel et al.(1998)]{1998ApJ...500..525S} Schlegel, D.~J., Finkbeiner, D.~P., \& Davis, M.\ 1998, \apj, 500, 525 
\bibitem[Toloba et al.(2009)]{2009ApJ...707L..17T} Toloba, E., Boselli, A., Gorgas, J., et al.\ 2009, \apjl, 707, L17 
\bibitem[Toloba et al.(2011)]{2011A&A...526A.114T} Toloba, E., Boselli, A., Cenarro, A.~J., et al.\ 2011, \aap, 526, A114 
\bibitem[Toloba et al.(2012)]{2012A&A...548A..78T} Toloba, E., Boselli, A., Peletier, R.~F., et al.\ 2012, \aap, 548, A78 
\bibitem[Toloba et al.(2015)]{2015ApJ...799..172T} Toloba, E., Guhathakurta, P., Boselli, A., et al.\ 2015, \apj, 799, 172 
\bibitem[Tonnesen \& Bryan(2009)]{2009ApJ...694..789T} Tonnesen, S., \& Bryan, G.~L.\ 2009, \apj, 694, 789 
\bibitem[Trentham \& Hodgkin(2002)]{2002MNRAS.333..423T} Trentham, N., \& Hodgkin, S.\ 2002, \mnras, 333, 423 
\bibitem[van Dokkum et al.(2015)]{2015ApJ...804L..26V} van Dokkum, P.~G., Romanowsky, A.~J., Abraham, R., et al.\ 2015, \apjl, 804, L26 
\bibitem[Voyer et al.(2014)]{2014A&A...569A.124V} Voyer, E.~N., Boselli, A., Boissier, S., et al.\ 2014, \aap, 569, AA124 
\bibitem[Whitmore et al.(1993)]{1993ApJ...407..489W} Whitmore, B.~C., Gilmore, D.~M., \& Jones, C.\ 1993, \apj, 407, 489 
\bibitem[Wright et al.(2010)]{2010AJ....140.1868W} Wright, E.~L., Eisenhardt, P.~R.~M., Mainzer, A.~K., et al.\ 2010, \aj, 140, 1868 
\bibitem[Wyder et al.(2005)]{2005ApJ...619L..15W} Wyder, T.~K., Treyer, M.~A., Milliard, B., et al.\ 2005, \apjl, 619, L15 
\bibitem[Zhang et al.(2015)]{2015ApJ...802...30Z} Zhang, H.-X., Peng, E.~W., C{\^o}t{\'e}, P., et al.\ 2015, \apj, 802, 30 



\end{thebibliography}
\end{document}